\documentclass{aa}
\usepackage{graphicx}
\usepackage{txfonts}
\usepackage{mhchem,booktabs}

\usepackage[pdftex,colorlinks=true]{hyperref}
\graphicspath{{./}{figures/}}
\usepackage{color}

\def\fcmaps{Fig.~\ref{fig:cmaps}}

\def\alico{\texttt{ALICO }}

\def\myddag{\ensuremath{\ddag}}
\def\mypm{\ensuremath{\pm}}
\def\tempin{\ensuremath{T_{\rm in}}}
\def\tempout{\ensuremath{T_{\rm out}}}

\def\hcn{\ensuremath{{\ce{HCN}}}}
\def\trhcn{\ensuremath{{\ce{H^{13}CN}}}}
\def\thcn{\ensuremath{{\ce{^{13}CN}}}}
\def\fihcn{\ensuremath{{\ce{HC^{15}N}}}}

\def\fifn{\ensuremath{\ce{^{15}N}}}

\def\twc{\ce{^{12}C}}
\def\thc{\ce{^{13}C}}

\def\myddag{\ensuremath{^{\ddag}}}

\def\mypm{\ensuremath{\pm}}

\newcommand{\kms}{\ensuremath{\rm\,km\,s^{-1}}}
\newcommand{\ccc}{\ensuremath{\rm\,cm^{-3}}}

\def\micr{\ensuremath{\,\mu\rm m}}

\def\tdust{\ensuremath{T_{\rm d}}}
\def\tempin{\ensuremath{T_{\rm in}}}
\def\tempout{\ensuremath{T_{\rm out}}}

\def\ccg{\ensuremath{\,\rm cm^2\,g^{-1}}}
\def\hh{\ce{H2}}

\def\nhh{\ensuremath{n_{\hh}}}

\newcommand{\cratio}[0]{\ensuremath{\mathrm{~^{12}C/^{13}C}}}
\newcommand{\nratio}[0]{\ensuremath{\mathrm{~^{14}N/^{15}N}}}
\newcommand{\errnratio}[2]{\ensuremath{\mathrm{~^{14}N/^{15}N=#1\pm#2}}}
\newcommand{\euplow}[2]{\ensuremath{^{+#1}_{-#2}}}

\newcommand{\jtr}[2]{\ensuremath{J={#1}\rightarrow{#2}}}
\newcommand{\ftr}[2]{\ensuremath{F={#1}\rightarrow{#2}}}
\newcommand{\jlev}[1]{\ensuremath{J={#1}}}

\newcommand{\tdix}[1]{\ensuremath{\times10^{#1}}}
\newcommand{\dix}[1]{\ensuremath{10^{#1}}}

\newcommand{\cm}[1]{\ensuremath{\mathrm{ cm}^{#1}}}
\newcommand{\tex}[0]{\ensuremath{T_{\mathrm{ ex}}}}

\begin{document}

\title{Abundance of HCN and its C and N isotopologues in L1498}

\author{%
  V. S. Magalh\~{a}es\inst{1}
  \and
  P. Hily-Blant\inst{1,2}
  \and
  A. Faure\inst{1}
  \and
  M. Hernandez-Vera\inst{3}
  \and
  F. Lique\inst{2,3}
  }

\institute{%
  Institut de Plan\'{e}tologie et d'Astrophysique de Grenoble,
  Universit\'{e} Grenoble Alpes,
  414 Rue de la Piscine, Grenoble CEDEX 9\\
  \email{victor.de-souza-magalhaes@univ-grenoble-alpes.fr} \and
  Institut Universitaire de France, 75231 Paris Cedex 05, France\and
  LOMC-UMR 6294, CNRS-Universit\'{e} du Havre, 25 rue Philippe Lebon, BP 540, 76058 Le Havre, France
}

\date{}

\abstract{%
  The isotopic ratio of nitrogen in nearby protoplanetary disks,
  recently measured in CN and HCN, indicates that a fractionated
  reservoir of volatile nitrogen is available at the earliest stage of
  comet formation. This reservoir also presents a 3:1 enrichment in
  $^{15}$N relative to the elemental ratio of 330, identical to that
  between the solar system comets and the protosun, suggesting that
  similar processes are responsible for the fractionation in the
  protosolar nebula (PSN) and in these PSN analogs. However, where,
  when, and how the fractionation of nitrogen takes place is an open
  question. Previously obtained HCN/HC$^{15}$N abundance ratios
  suggest that HCN may already be enriched in $^{15}$N in prestellar
  cores, although doubts remain on these measurements, which rely on
  the double-isotopologue method. Here we present direct measurements
  of the HCN/H$^{13}$CN and HCN/HC$^{15}$N abundance ratios in the
  L1498 prestellar core based on spatially resolved spectra of
  HCN(1-0), (3-2), H$^{13}$CN(1-0), and HC$^{15}$N(1-0) rotational
  lines. We use state-of-the-art radiative transfer calculations using
  \texttt{ALICO}, a 1D radiative transfer code capable of treating
  hyperfine overlaps. From a multiwavelength analysis of dust emission
  maps of L1498, we derive a new physical structure of the L1498
  cloud.  We also use new, high-accuracy HCN-H$_2$ hyperfine
  collisional rates, which enable us to quantitatively reproduce all
  the features seen in the line profiles of HCN(1-0) and HCN(3-2),
  especially the anomalous hyperfine line ratios. Special attention is
  devoted to derive meaningful uncertainties on the abundance
  ratios. The obtained values, HCN/H$^{13}$CN=45$\pm$3 and
  HCN/HC$^{15}$N=338$\pm$28, indicate that carbon is heavily
  fractionated in HCN, but nitrogen is not. For the
  H$^{13}$CN/HC$^{15}$N abundance ratio, our detailed study validates
  to some extent analyses based on the single excitation temperature
  assumption. Comparisons with other measurements from the
  literature suggest significant core-to-core
  variability. Furthermore, the heavy $^{13}$C enrichment we found in
  HCN could explain the superfractionation of nitrogen measured in
  solar system chondrites.}


\maketitle

\def\refdust{Table~\ref{tab:cresults}}

\section{Introduction}

The link between the chemical composition of the protosolar nebula
(PSN) and of the interstellar medium (ISM) may be more direct than
previously thought. Observations of \ce{O2} and \ce{S2} in the coma of
comet 67P/Churyumov-Gerasimenko performed with the ROSINA instrument
on board the ESA/\emph{ROSETTA} spacecraft (\citealt{leroy2015} and
\citealt{bieler2015}) indicate that cometary ices are, to some extent,
of interstellar nature (\citealt{taquet2016}, \citealt{calmonte2016}
and \citealt{mousis2017}).  An interstellar origin for cometary ices
is further supported by observations of water toward protostars,
indicating that only a minor amount (less than 10--20\% in abundance)
of water ices indeed sublimate at this stage \citep{vandishoeck2014a},
thus leaving substantial amounts of pristine interstellar material to
build up cometary ices. Nevertheless, to which extent comets have
preserved the composition of the parent interstellar cloud remains an
open question. One outstanding problem is that of the origin of
nitrogen in comets.

Unlike the deuterium-to-hydrogen isotopic ratio, the \nratio\ isotopic
ratio in solar system comets is strikingly uniform, with an average
\nratio=144$\pm3$, independent of the molecular carrier (\ce{NH2},
\ce{CN}, \hcn) and comet family (\citealt{mumma2011},
\citealt{shinnaka2016} and \citealt{hilyblant2017}, hereafter
HB17). This low value, three times lower than the elemental isotopic
ratio in the PSN, \nratio=441$\pm$5 as traced by the protosun and
Jupiter \citep{marty2011}, indicates that comets carry a fractionated
reservoir of nitrogen. The central question that motivates the present
work is to know when, where, and how this reservoir was formed.

The fractionated reservoir observed in comets may have three different
origins: \textit{i)} inherited from the parent interstellar cloud,
\textit{ii)} built in situ in the PSN at the epoch of comet formation,
\textit{iii)} or within the comets themselves over the last
4.6~Gyr. Recent progress has shown that the third possibility is not
necessary, while theoretical and observational difficulties preclude a
choice to be made between the interstellar and PSN scenarios.

A PSN origin is partially supported by models of selective
photodissociation of molecular nitrogen, \ce{N2}, showing that the
radiative and thermal conditions prevailing in protoplanetary disks
are prone to an enrichment in \fifn\ of several species, such as HCN
and CN. The enrichment predicted for HCN is indeed in good agreement
with the isotopic ratio measured indirectly in disks
\citep{guzman2017}, which yield an average of 111$\pm19$ (HB17)
although large uncertainties of up to 100\% on the assumed HCN/H\thcn\
ratio cannot be ruled out. Nevertheless, selective photodissociation
models also predict significant enrichments in CN, which is in
disagreement with the directly measured CN/C\fifn\ ratio of 323$\pm$30
in the TW~Hya disk (HB17). HB17 argued that CN is not fractionated,
and that the CN/C\fifn\ ratio reflects the present-day elemental ratio
in the local ISM, for which we adopt an average value of $\approx 330$
with a typical uncertainty of 10\%. The HCN/HC\fifn\ ratio in PSN
analogs is thus three times lower than the bulk, which suggests that
the fractionated reservoir of nitrogen recorded by comets is already
available at the earliest stages of planet formation, thus making
fractionation within comets over the last 4.6 Gyr unnecessary. We note
that the value of 330 is higher than, but consistent with, previous
estimates, in particular the value of 290$\pm$40 derived from an
interpolation of the galactic gradient of \nratio\ \citep{adande2012},
but is in remarkable agreement with the predictions from most recent
Galactic evolution models \citep{romano2017}. In the PSN scenario, the
similar threefold enrichment in HCN in disks with respect to the bulk
therefore suggests that the same fractionation process shapes the
nitrogen isotopic ratio in these disks and in the PSN at the epoch of
comet formation. Nevertheless, the sample of disks surveyed in
\cite{guzman2017} encompasses a broad range of masses and ages, and it
remains to be demonstrated that selective photodissociation in PSN
analogs can lead to similar enrichments in \fifn\ despite these
different irradiation fields and dust size distributions.

In the interstellar origin scenario, one would expect to find
  \fifn-rich isotopic ratios typically three times lower than the bulk
  in contracting clouds, as early as the prestellar core stage or
  later, in protostars. The picture is confusing, however,
  because on the one hand, some observations of prestellar cores
  evince the expected fractionated reservoir of nitrogen
  \citep{hilyblant2013icarus}. On the other hand, the most recent
  chemical models (\citealt{roueff2015} and \citealt{wirstrom2018})
  predict instead that chemical mass fractionation is essentially
  inefficient in cold clouds. There may be issues on both the
  theoretical and the observational sides, but it is true that
  measuring accurate isotopic ratios in cold clouds (and also in
  protoplanetary disks) is a challenging task.  In particular, it is
important to distinguish between direct measurements (using the main
isotopologue, e.g., CN/C\fifn) from indirect measurements based on
double-isotopic ratios (e.g., \trhcn/HC\fifn). Chemical models
  incorporating both carbon and nitrogen fractionation have stressed
  the caveats of the double-isotopic method \citep{roueff2015}. In
prestellar cores, direct measurements are scarce. In hydrides \ce{
  (NH3} and \ce{NH2D}) and in \ce{N2H+}, a daughter species of
\ce{N2}, directly obtained ratios yield a weighted average of
$336\pm16$, thus indicating that these species are not
fractionated. For nitriles (\ce{CN}, \hcn, and \ce{HNC}, which are
daughter molecules of atomic nitrogen) ratios are usually obtained
indirectly (\citealt{ikeda2002} and \citealt{hilyblant2013icarus,
  hilyblant2013c15n}). In some instances, however, hyperfine splitting
of the rotational lines may provide optically thin transitions, as in
CN, which can be used to directly infer the total column density
\citep{adande2012}, although departures of the hyperfine intensity
ratios from single excitation temperature predictions can perturb the
analysis. For nitriles such as \ce{HC3N} and \ce{HC5N}, abundances are
intrisically low and the \nratio\ ratio can be measured directly
(\citealt{taniguchi2017a} and
\citealt{hilyblant2018a}). Interestingly, the \ce{HC5N}/\ce{HC5^{15}N}
abundance ratio was measured both directly and indirectly toward the
cyanopolyyne peak of TMC-1, leading to \errnratio{344}{53} and
\errnratio{323}{80,} respectively, in harmony with the elemental ratio
of 330. For \ce{HC3N}, the indirect and direct methods also agree
within 1$\sigma$, and the derived average value is 264$\pm$40. This
ratio is only marginally compatible with the local ISM, suggesting a
secondary reservoir of nitrogen in the ISM. For the
double-isotopologue method, the values of \cratio\ were measured for
all carbon atoms in either \ce{HC3N} or \ce{HC5N}
\citep{taniguchi2016a}, making the determination more robust than
would be otherwise expected. However, as pointed out recently in the
case of \ce{HC3N} in the \object{L1544} prestellar core, different
analyses of the same data can lead to ratios that may differ by almost
a factor of 2 \citep{hilyblant2018a}.

\begin{figure}
\centering
 \includegraphics[width=\hsize]{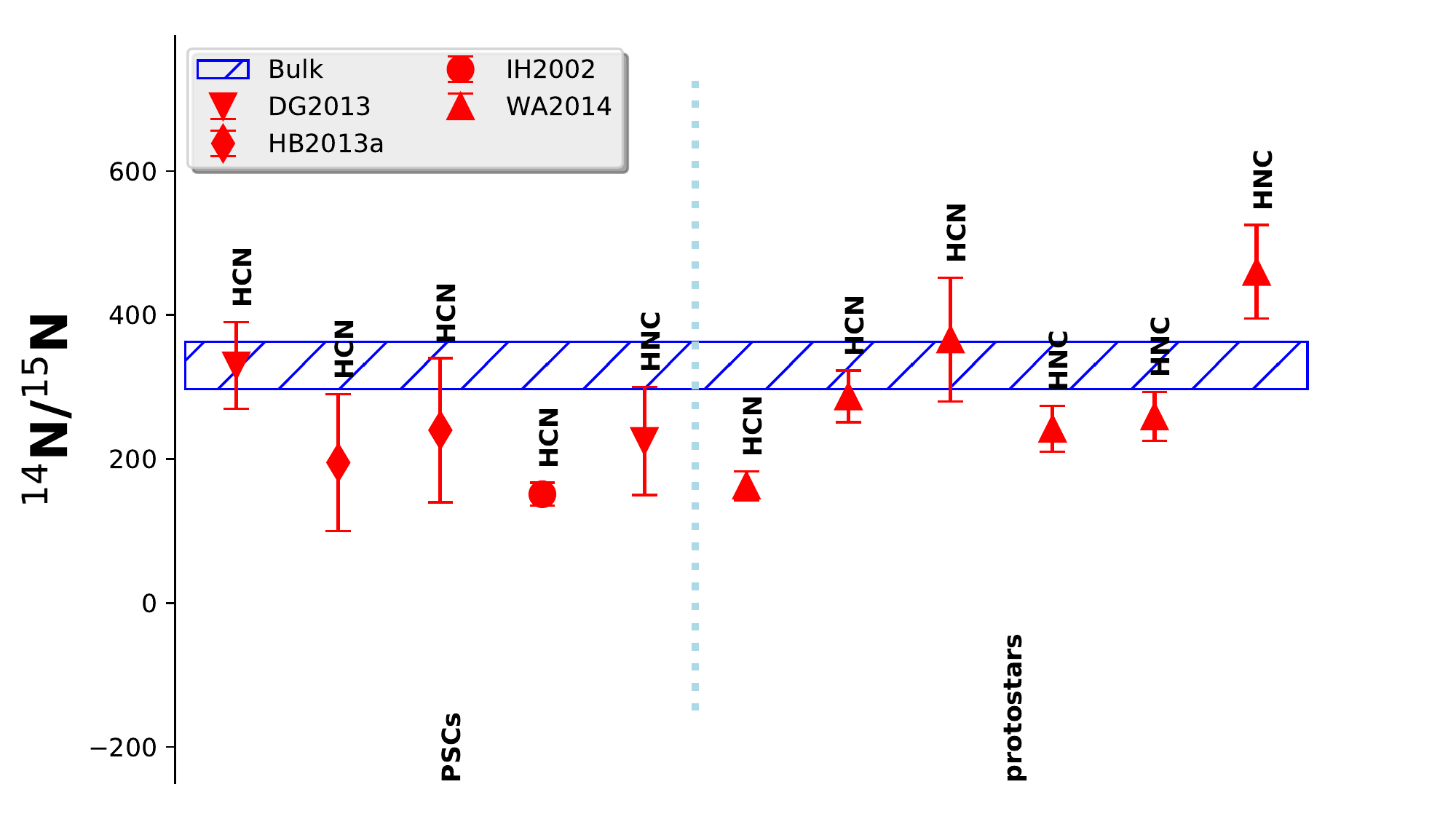}
 \caption{Published nitrogen isotopic ratios for HCN and HNC toward
   prestellar cores (PSCs) and protostellar envelopes. DG2013 is
   \cite{daniel2013}, WA2014 is \cite{wampfler2014}, IH2002 is
   \cite{ikeda2002} and HB2013a is \cite{hilyblant2013icarus}.}
\label{fig:hcn_hnc_nratio}
\end{figure}

Isotopic ratios obtained in HCN and HNC toward prestellar and
protostellar cores are summarized in Fig.~\ref{fig:hcn_hnc_nratio}.
All the reported measurements are indirect and were obtained adopting
HCN/\trhcn= 70, except for \cite{ikeda2002}, where HCN/\trhcn=90.
Attempts to obtain direct isotopic ratios in the evolved B1 collapsing
core were presented in \cite{daniel2013} but were discarded by the authors
themselves, and are thus not reported. As is evident from
Fig.~\ref{fig:hcn_hnc_nratio}, these measurements show a large scatter,
but we note that they almost all remain within the lower and upper
bounds corresponding to the fractionated reservoir in protoplanetary
disks (111$\pm$19) and to the elemental ratio of 330,
respectively. Nevertheless, direct isotopic ratio measurements are
needed to confirm the interstellar origin of the fractionated
reservoir in disks.

The chief objective of the present work is the direct measurement of
the nitrogen isotopic ratio in \hcn\ in the \object{L1498} prestellar
core, at the southwest end of the Taurus-Auriga complex, located in
the solar neighborhood, at a distance of 140~pc from the Sun. To
achieve our aims, we use the emission spectra of the \jtr{1}{0} and
\jtr{3}{2} rotational transitions of HCN and the \jtr{1}{0} transition
of \trhcn\ and \fihcn, along a cut through L1498
(Section~\ref{sec:obs}). Spectra generated using our accelerated
lambda iteration code, \alico\ \citep{daniel2008}, are minimized
against the observed spectra to infer the abundances of the three
isotopologues (Section~\ref{sec:lmodel}).  Our model shares
similarities with previous analyses of the HCN(1-0) spectra from
contracting starless cores similar to L1498 \citep{lee2007}. However,
the present study includes lines of HCN isotopologues and of HCN(3-2)
that bring additional, very sensitive constraints, especially on the
velocity field inside the core. In the process, a physical model of
the source was derived by a simultaneous fit to Herschel/SPIRE and
IRAM/MAMBO continuum maps at 250, 350, 500, and 1250\micr\
(Section~\ref{sec:cmodel}).

\begin{table*}[t]
  \centering
  \caption{\label{tab:obs}Summary of the line observations. The
    reference position for L1498 is
    $(\alpha,\delta)_{J2000}$=04:10:52.2, +25:10:20.}
  \begin{tabular}{l ccccc}
    \toprule
    Line & Rest. Freq. & $\delta v$ 
    & $T_{\rm sys}$ & rms  & Program \\
         & MHz & \kms & K & mK [Tmb] \\
    \midrule
    HCN(1-0)    & 88631.6022 & 0.066  & 115& 50 & 031-11 \\
                &            & 0.066 &   100& 20 & 008-16 \\
    HCN(3-2)    &265886.4339 & 0.022   & 170& 36 & 105-16 \\
    \trhcn(1-0) & 86339.9214 & 0.068   & 90 & 10 & 007-16 \\
    \fihcn(1-0) & 86054.9664 & 0.068   & 90 & 10 & 007-16 \\
    \bottomrule
  \end{tabular}
\end{table*}

\def\ww{0.23\hsize}
\begin{figure*}
  \centering
  \includegraphics[height=\ww]{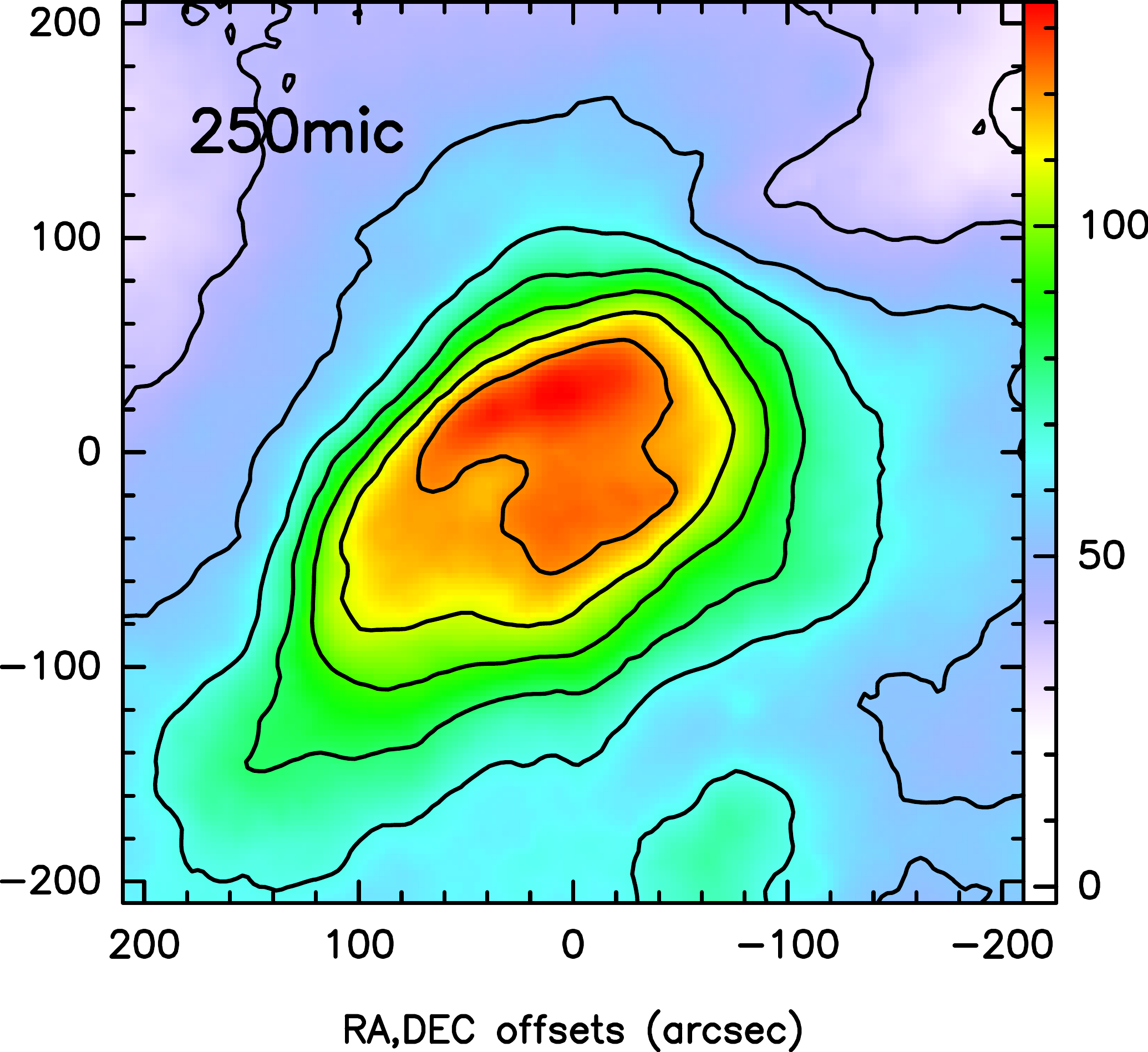}\hfill% \hskip10ex
  \includegraphics[height=\ww]{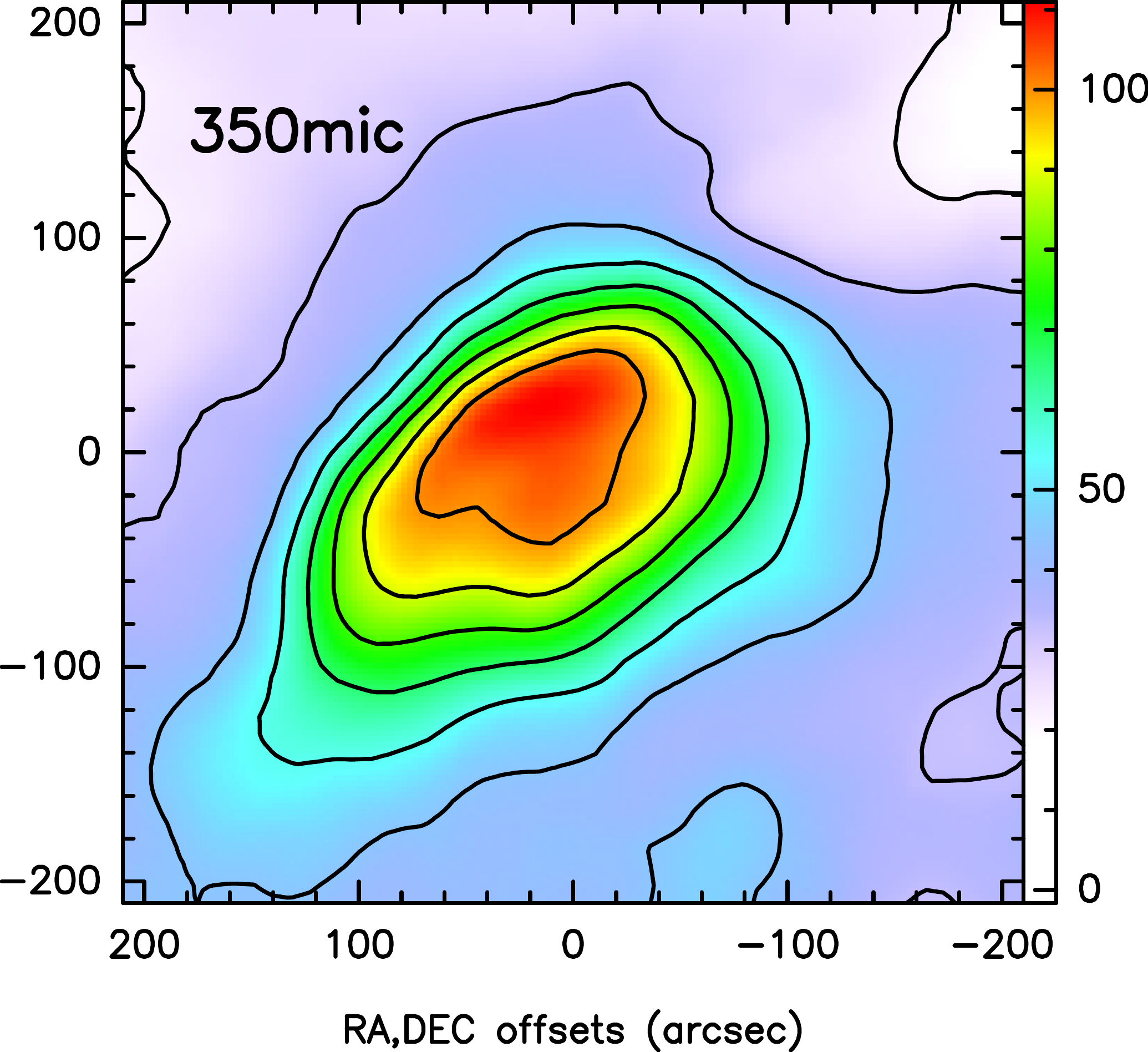}\hfill%bigskip\\
  \includegraphics[height=\ww]{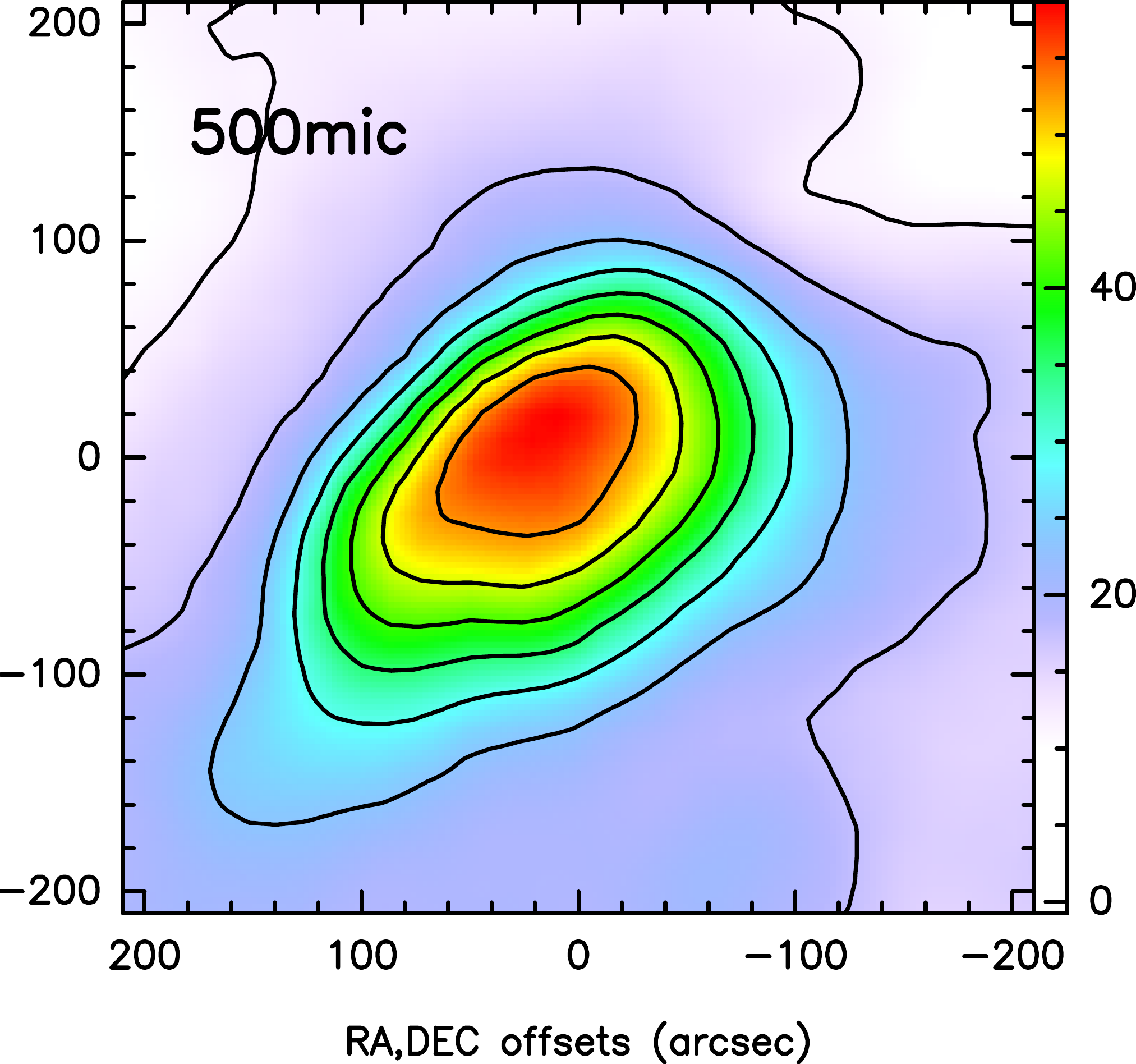}\hfill%\hskip11ex
  \includegraphics[height=\ww]{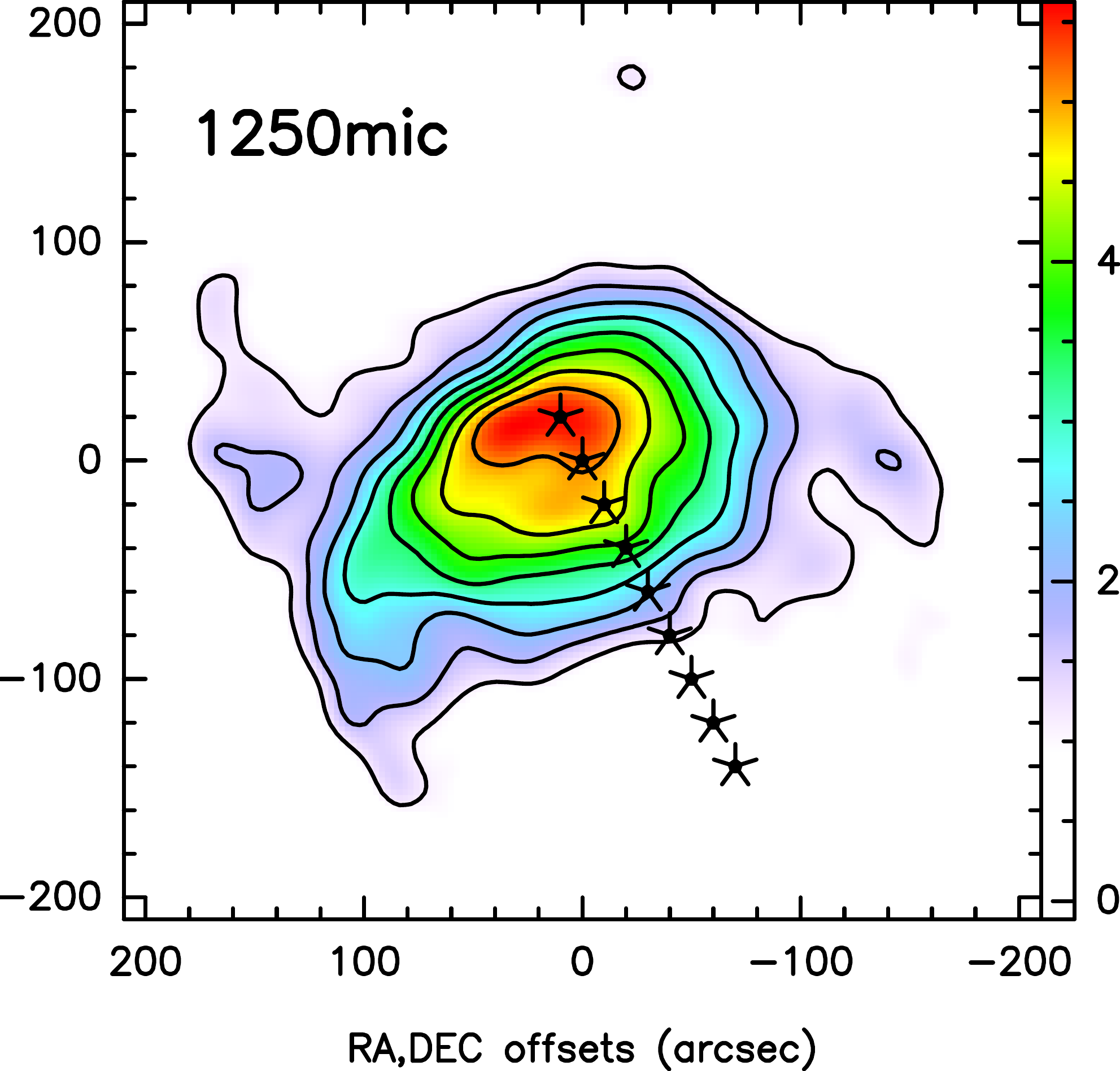}
  \caption{Herschel/SPIRE and IRAM/MAMBO maps of the continuum
    emission (in MJy/sr) at 250, 350, 500, and 1250\micr. The HPBW is
    18, 24, 35, and 11\arcsec\ respectively. Contours are evenly
    spaced between 20 to 90\% of the maximum intensity. The reference
    position is $(\alpha,\delta)_{J2000}$=04:10:52.2,
    +25:10:20. 
    }
  \label{fig:cmaps}
\end{figure*}

\section{Observations}
\label{sec:obs}

Our observations can be divided into two datasets: spectra of HCN and
its isotopologues, used to infer the abundances of HCN and its
isotopologues; and a set of continuum maps used to derived the radial
density profile of L1498.

\subsection{HCN and its isotopologues}

All our observations of \hcn\ and its isotopologues were obtained at
the IRAM-30m telescope, but in different runs (see Table~\ref{tab:obs}
for a summary).  All spectra have been reduced and analyzed using the
IRAM GILDAS/CLASS software. Bandpass calibration was measured every 15
min using the standard three-steps method implemented at IRAM-30m. The
typical receiver and system temperatures are 100 and 170~K at 89 and
266~GHz, respectively. The obtained rms (main-beam temperature scale)
is 10~mK for the \trhcn\ and \fihcn(1-0), 20 to 50~mK for HCN(1-0),
and 36~mK for HCN(3-2). Pointing and focus were checked regularly on a
basis of one to two hours, leading to a pointing accuracy of typically
1-2\arcsec\ at all frequencies. Finally, residual bandpass effects
were removed by subtraction of low-order polynomials (up to degree
3). Table~\ref{tab:obs} summarizes the observation setups and
performances. Spectra were brought into a main-beam temperature scale
using tabulated beam efficiencies appropriate at the time of
observations. Figures~\ref{fig:obshcn} and \ref{fig:obsiso} show the
final set of spectra.

\subsection{Continuum}
\label{sec2:continuum}

We retrieved the Herschel/SPIRE maps at 250, 350 and 500 $\mu m$ from
the Gould Belt Survey \citep{andre2010} available from the Herschel
public archive. We used Level 3 data, which were calibrated following
the standard \emph{extended source} procedure of pipeline SPG
v14.1.0. In the process, an offset was added to the images to comply
with ESA/Planck absolute intensities. The continuum map at 1.25~mm was
obtained at the IRAM-30m telescope with the MAMBO 37-channels bolometer
array \citep{tafalla2002}. The continuum maps (in MJy/sr) are
shown in \fcmaps. For the Herschel maps, the statistical noise was computed
as the quadratic sum of the calibration noise (as estimated by the
pipeline and taken from the file header) and the contribution from the
spatially varying contribution from cirrus-like emission, which
shows up
in histograms of the intensity as a broad, low-level
Gaussian. Two-Gaussian fits to these histograms recover
the calibration uncertainty well. At 1250\micr, the noise was estimated in
a similar fashion, although a single Gaussian was sufficient. The final
noise is 4.1, 2.2, 1.0, and 0.6 MJy/sr at 250, 350, 500, and
1250\micr\ respectively.

\def\ww{0.3\textheight}
\begin{figure*}
  \centering
  \includegraphics[height=\ww]{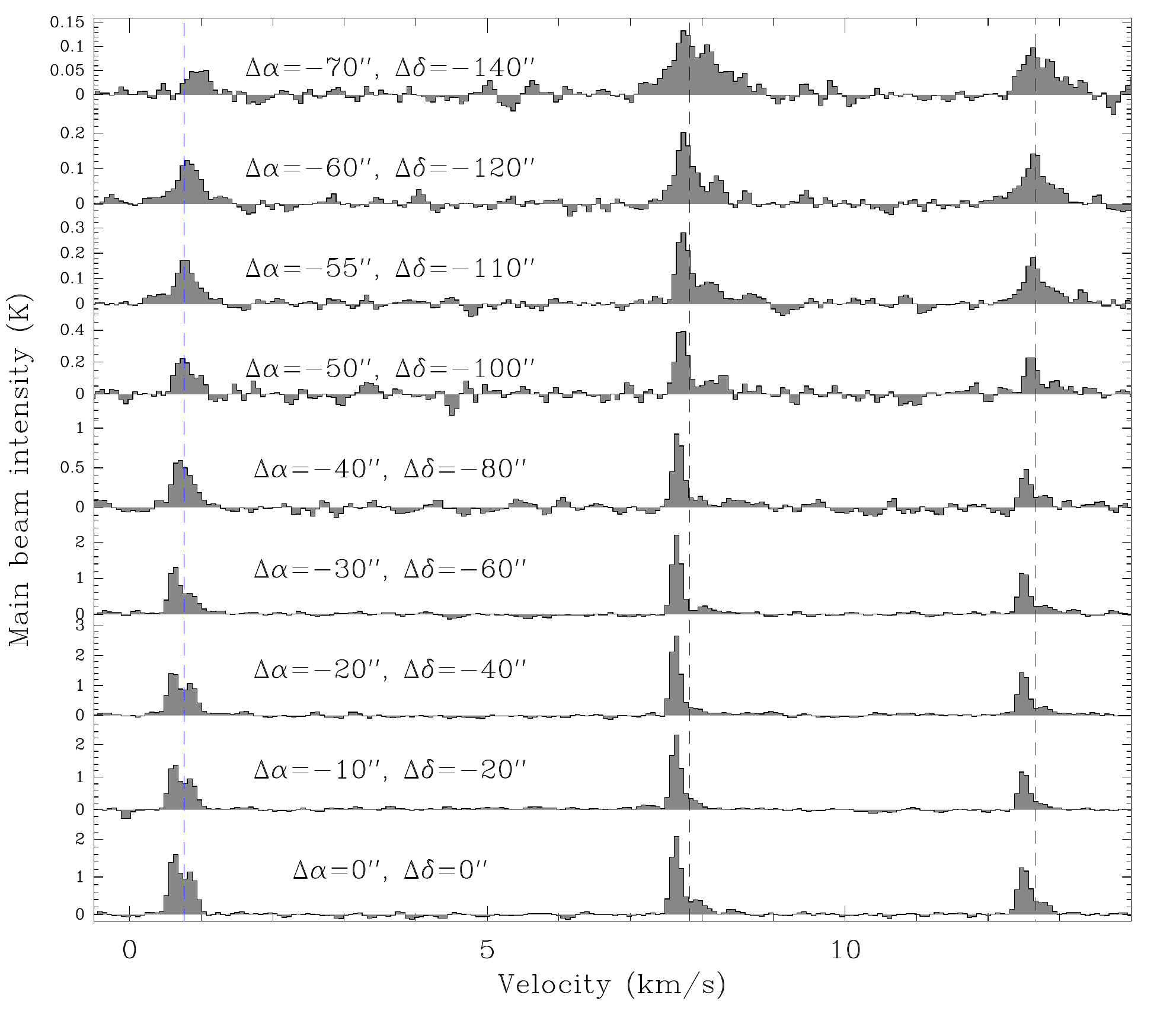}%
  \hfill%
  \includegraphics[height=\ww]{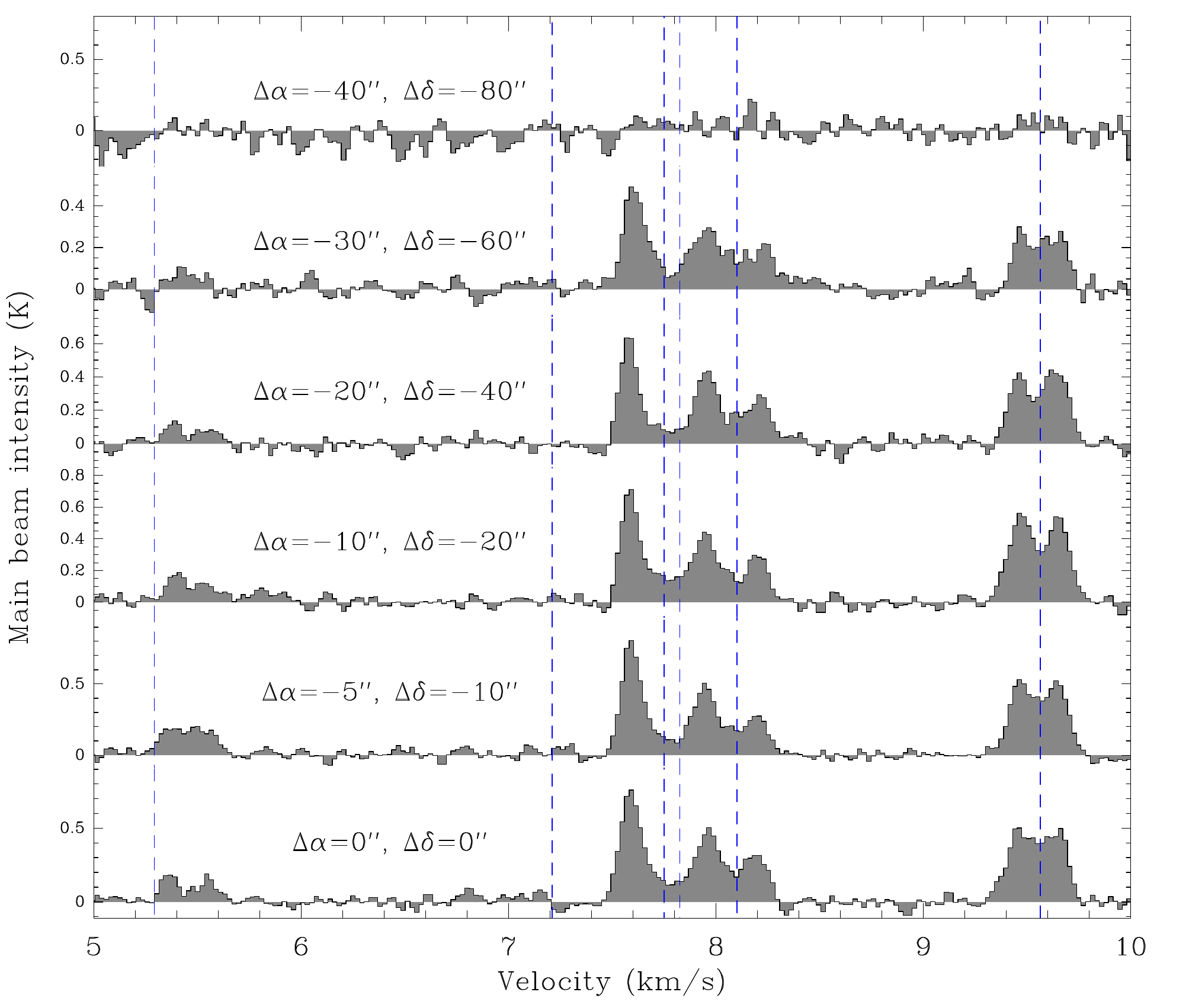}
  \caption{\hcn\jtr{1}{0} (left) and \jtr{3}{2} (right) spectra at
    several positions toward L1498 (RA-DEC offsets from the continuum
    peak position ($\alpha$ = 4:10:52.2, $\delta$ = 25:10:20, J2000) are in seconds of arc). The systemic velocity for
    each of the hyperfine components is indicated (dashed blue
    lines).}
  \label{fig:obshcn}
\end{figure*}

\begin{figure*}
  \centering
  \includegraphics[height=\ww]{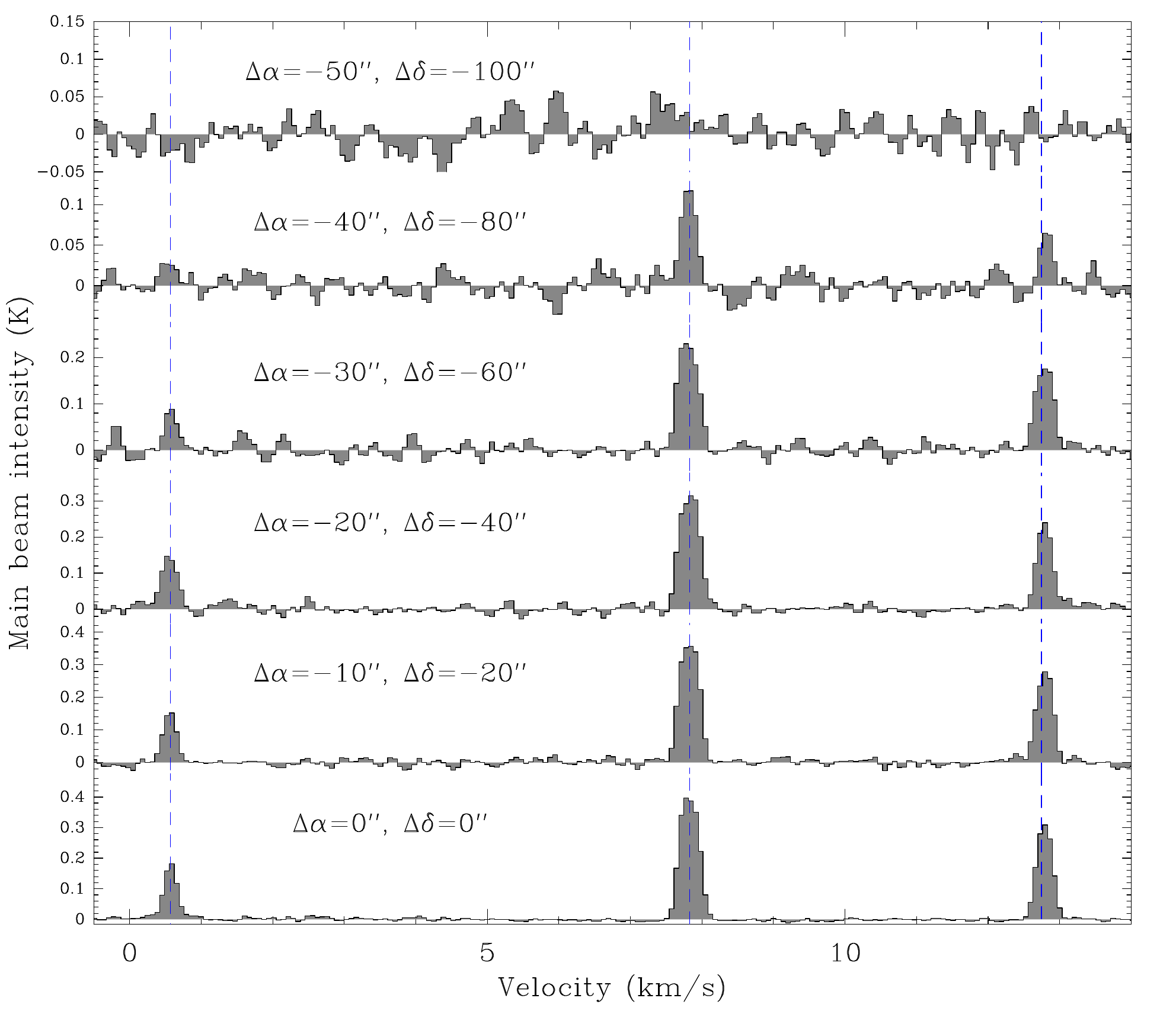}%
  \hfill%
  \includegraphics[height=\ww]{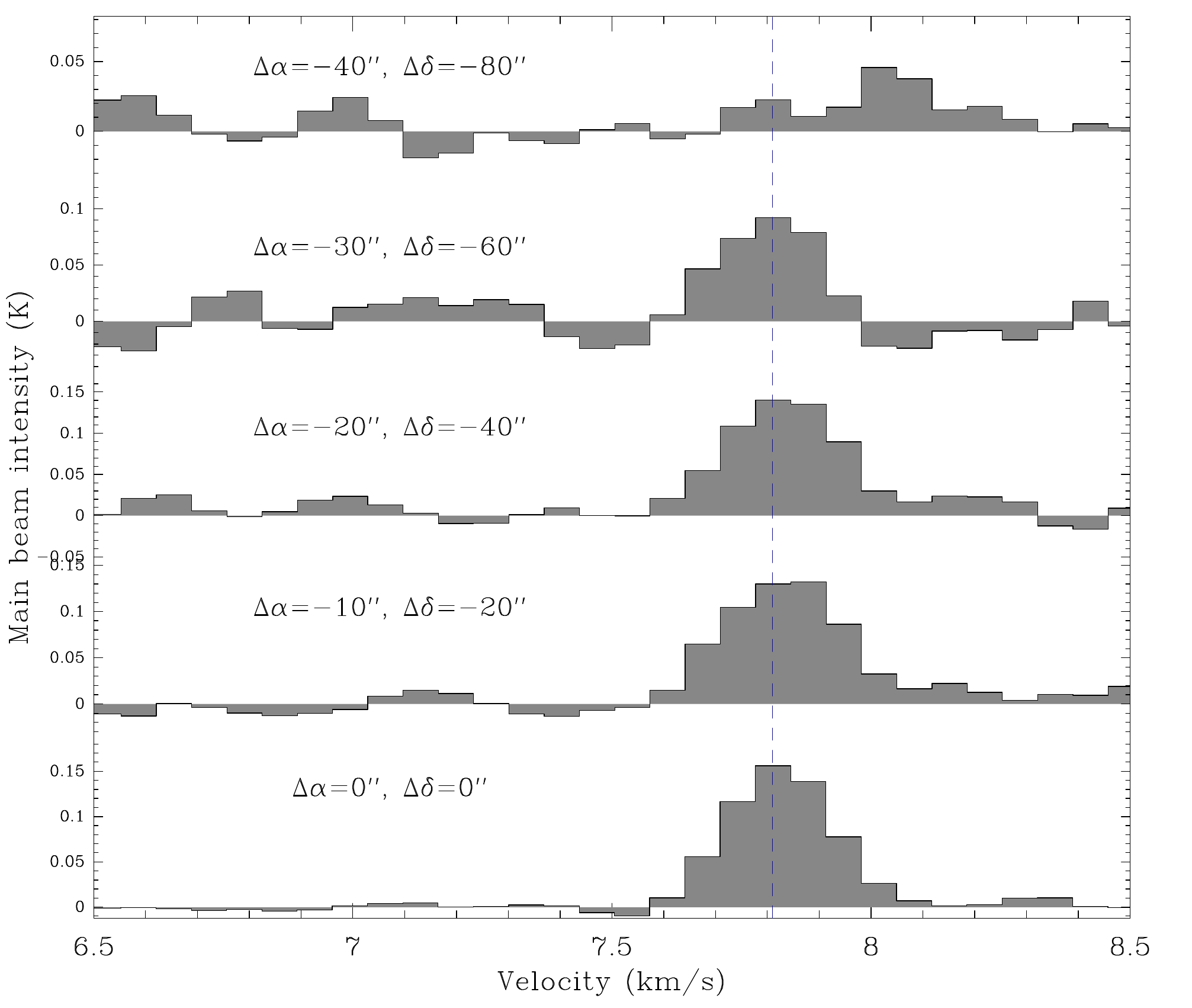}
  \caption{Same as Fig.~\ref{fig:obshcn} for the \jtr{1}{0} transition
    of \trhcn\ (left) and \fihcn\ (right).}
  \label{fig:obsiso}
\end{figure*}

\section{Physical model of L1498 from the dust emission}
\label{sec:cmodel}

The line radiative transfer calculations of the lines of HCN and its
isotopologues toward L1498 compute the populations at the hyperfine
level in spherical geometry, assuming density, temperature, and
velocity profiles, as well as a profile of the velocity
dispersion. The number of free parameters can become outrageously
large in such simulations, but in principle, all these parameters could
be minimized (at least after adopting some simple parameterization
for each) using the various transitions at the several locations
along the cut. Nevertheless, we have used maps of the dust-dominated
continuum emission toward L1498 (\fcmaps) to obtain independent
constraints on the density profile, which is critical to the
calculation of the molecular excitation. The present analysis
encompasses and extends beyond earlier works \citep{tafalla2006} by
considering the dust emission at four wavelengths, which allows us to
derive constraints on the density and dust temperature profiles. In
the process, the spectral index is also measured, and was found in very
good agreement with comprehensive studies. We here review the basic
model and assumptions used in the analysis of the dust emission maps.

\begin{figure}
  \centering
  \includegraphics[width=\hsize]{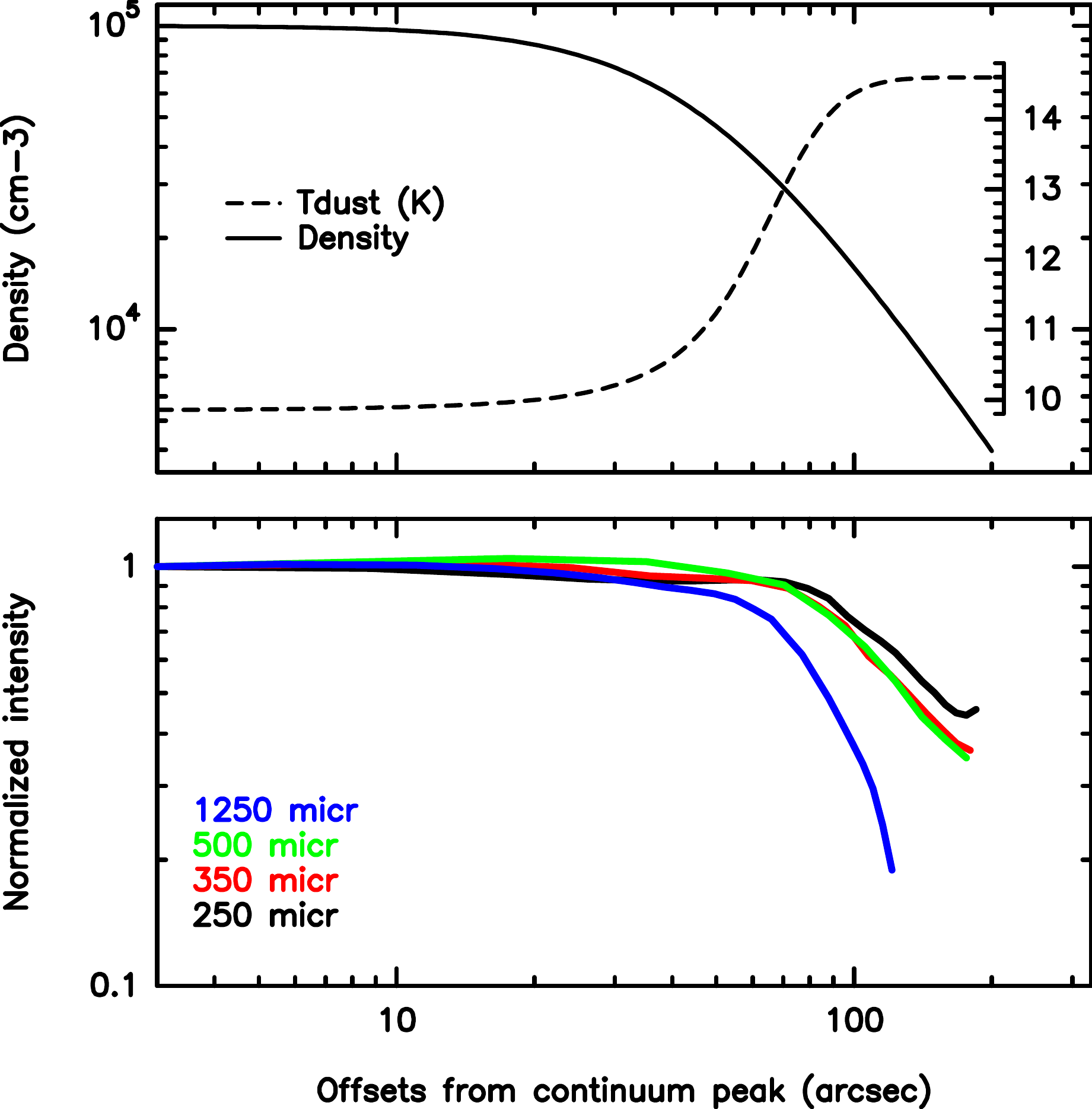}
  \caption{Physical structure adopted from the continuum map
    fitting. Top panel: Density and dust temperature (inset scale)
    profiles. Middle panel: Normalized continuum intensity at 250,
    350, 500, and 1250\micr\ along the HCN cut (see \fcmaps).
    }
  \label{fig:structure}
\end{figure}

\subsection{Dust emission}

From Kirchoff's law, the emissivity of the dust is directly related
to the absorption coefficient, noted $\kappa_\nu$ (in \ccg) at
frequency $\nu$, which we hereafter express by gram of dust and gas,
assuming a standard gas-to-dust mass ratio of 1\%. The modeling of the
dust emission then follows a standard approach \citep{bergin2007} in
which the optically thin dust emission is treated as a gray-body
radiation,
\begin{equation}
  \label{eq:inudust}
  I_\nu = 2 \int_0^X \kappa_\nu  \nhh(x) \mu_{\ce{H2}} m_{\ce{H}} B_\nu[\tdust(x)] dx
,\end{equation}
where the integral extends over the entire line of sight.  In our
notation, $m_{\ce{H}}$ is the mass of one hydrogen atom,
$\mu_{\ce{H2}}=2.8$ is the mean molecular weight per \ce{H2} molecule,
assuming 10\% of helium atoms with respect to total hydrogen
nuclei. $B_\nu(\tdust)$ is the emission of a blackbody at temperature
\tdust\ and frequency $\nu$. We note that in our formulation,
$\kappa_\nu$ is assumed to be uniform, but the dust temperature is
allowed to vary along the line of sight.

The frequency dependence of the dust absorption coefficient is
described by a power law,
\begin{equation}
  \kappa_\nu = \kappa_{250} (\lambda/250\micr)^{-\beta}
  \label{eq:kappa}
  .\end{equation}
This parameterization appears to hold reasonably well at the submillimeter to
millimeter wavelengths of concern in our analysis, although the value of
$\kappa_{250}$ and of the spectral index $\beta$ are known to vary with
the temperature and composition of the dust particles
(\citealt{juvela2015b} and \citealt{demyk2017}). With such a parameterization of
$\kappa_\nu$, $\beta$ primarily constrains the long wavelength,
Rayleigh-Jeans domain of the emission where
$B_\nu \propto \nu^{2+\beta}$. On the other hand, the value of
$\kappa_{250}$ fixes the dust column density, but not the shape of the
spectral energy distribution (SED). Typical values of $\beta$ that are appropriate for dense gas are
in the range 1.5--3, with $\beta=2$ being a commonly adopted value
that describes the asymptotic behavior of $\kappa_\nu$ at long
wavelengths. Comprehensive multiwavelength analysis of dust emission
from cold, starless cores find $\beta\approx 1.9\pm0.3$ for dust
temperatures $\approx16\pm2$~K \citep[][Fig. 27]{juvela2015b}.

\begin{table}
  \begin{center}
    \caption{\label{tab:kappa}Compilation of values of the mass
      absorption coefficient $\kappa_\nu$ (in \ccg, expressed per gram
      of gas and dust) used in dense core studies. The values assume
      $\beta=2$ (see Eq.~\ref{eq:kappa}). Numbers in parentheses are
      powers of ten.}
    \begin{tabular}{ccccr}
      \toprule
      250\micr & 350\micr & 500\micr & 1250\micr & Reference\\
      \midrule
      0.10        & 5.1(-2) & 2.5(-2) &4.0(-3) & (1) \\
      0.125       & 6.4(-2) & 3.1(-2) &5.0(-3) & (2) \\
      0.14        & 7.1(-2) & 3.5(-2) &5.6(-3) & (3)\\
      0.09        & 4.7(-2) & 2.3(-2) &3.7(-3) & (4)$^\dag$\\
      0.05        & 2.6(-2) & 1.3(-2) &2.0(-3) & (5)$^\ddag$\\
      0.10        & 5.2(-2) & 2.6(-2) &4.0(-3) & (5)$^\ddag$\\
      0.2$\pm$0.1 & 1.0(-1) & 5.0(-2) &8.0(-3) & (6)\\
      0.2$\pm$0.1 & 9.2(-2) & 4.1(-2) &4.9(-3) & (6)$^\S$\\
      0.09        & 5.5(-2) & 3.1(-2) &7.5(-3) & This work$^{||}$\\
      \bottomrule
    \end{tabular}
  \end{center}
  {\it References:} (1) \cite{hildebrand1983} (2) \cite{tafalla2004}
  (3) \cite{roy2014} (4) \cite{juvela2015a} (5) \cite{demyk2017} (6)
  \cite{chacon-tanarro2017}. \textit{Notes:} $\dag$ Using their relation
  $\tau(250)=2.16\tdix{-25} N_{\rm H} \,\rm cm^2/H$ and assuming 10\%
  of He atoms with respect to hydrogen nuclei. $\ddag$ Values are
  from their model calculations (their Table~1); values from the
  synthesized samples, which are probably more emissive than
  interstellar dust grains, are higher by factors of 4 to 10. $\S$
  Using $\beta=2.3$ as obtained from their study of the dust emission
  at 1.2 and 2~mm.  $||$ Using $\beta=1.56$ as obtained from our continuum fit.
\end{table}

In our analysis we have adopted $\kappa_{250}=0.09\ccg$, which appears well suited
to our study of a dense starless core (see Table~\ref{tab:kappa}). The
spectral index $\beta$ was also assumed uniform, but was a free
parameter in the minimization process.

\subsection{Density profile}
\label{sec:densprof}

The density profile was parameterized in a form that describes an
inner plateau of almost constant density surrounded by a power-law
envelope (\citealt{bacmann2000} and \citealt{tafalla2002}):
\begin{equation}
  \label{eq:densprof}
  \nhh(r) = \frac{n_0}{1+(r/r_0)^\alpha}
  ,\end{equation}
where $n_{\ce{H2}}(r)$ is the \ce{H2} number density at radius $r$. The plateau
has a radius $r_0$ and a constant \ce{H2} density $n_0$, while the
density in the envelope decreases as $r^{-\alpha}$. This type of
profile is well suited to describe starless cores in which the
pressure gradient evolves on timescales longer than the sound
crossing-time, such that these cores evolve quasi-statically toward
collapse, well before the formation of the first hydrostatic core
(\citealt{larson1969} and \citealt{lesaffre2005}). As we describe later, L1498 is
precisely in such a state. This profile was adopted in
previous studies of the chemical structure of L1498
\citep{tafalla2006}. This density profile eventually merges with the
ambient medium with typical \ce{H2} density 500~\cm{-3}
(\citealt{hilyblant2007} and  \citealt{lippok2016}). However, with the typical
values of
$r_0$ and $n_0$ that describe L1498, such a density is obtained at
large radii (beyond 200\arcsec) and need not be considered in our dust
emission fitting (see Fig.~\ref{fig:structure}).

\subsection{Dust temperature profile}

The gas in a prestellar core such as L1498 with a relatively low
central density ($\approx \dix{5}$~\cm{-3}) is, to a good
approximation, constant and close to 10~K as long as CO cooling is
sufficient (\citealt{zucconi2001} and \citealt{tafalla2006}).  In the
inner parts where the density is high, collisions become efficient in
thermalizing the gas and the dust to temperatures close to and below
10~K \citep{makiwa2016}. However, the dust temperature is known to
increase outward from typically below 10~K to approximately 15~K or
more (\citealt{lippok2016} and \citealt{bracco2017}). Our attempts to
simultaneously fit the dust emission at the four wavelengths with a
constant dust temperature did not converge. To allow for a radial
increase, we have adopted a simple radial profile of the form
\begin{equation}
  \tdust(r) = \tempin + 
  \frac{\tempout-\tempin}{2}\,\left(1+\tanh\frac{r-r_d}{\Delta r_d}\right)
  \label{eq:tdust}
,\end{equation}
which describes a continuous increase from \tempin\ to \tempout, at a radius
$r_d$ and over a characteristic scale $\Delta r_d$. In the following,
\tempin, \tempout, $r_d$, and $\Delta r_d$ are considered free
parameters. Having a non-uniform dust temperature but uniform spectral
index $\beta$ may seem contradictory, as variations of $\beta$ with
\tdust\ have been reported in starless cores \citep{juvela2015b}. Our study is focused on the determination of the density profile
in view of an accurate measurement of the abundances of HCN and
isotopologues, however. We therefore decided to keep the number of free
parameters as low as possible. Moreover, the variations of $\beta$
with \tdust\ still remain elusive \citep{bracco2017}.

\subsection{Results}

With the above assumptions on the dust emissivity and on the adopted
density and dust temperature profiles, we performed a minimization
calculation using a Markov chain Monte Carlo (MCMC) approach to
explore the eight-dimension parameter space
(Table~\ref{tab:cresults}). The fitting of the gray-body emission was
calculated in a spherical geometry by integrating along lines of sight
at offsets evenly spaced along the southwest cut from the continuum
peak, where spectral line integrations have been performed (see
\fcmaps\ and the bottom panel of Fig.~\ref{fig:structure}). Our
calculation is thus similar to those of \citet{roy2014} and
\citet{bracco2017}. However, we did not attempt to compute the
derivative of the intensity to obtain $\nhh B_\nu$, but instead
adopted a forward approach in which the various parameters (density,
temperature, and $\beta$) are free to vary within predefined intervals
(with uniform probability) and the emerging intensity was computed
assuming optically thin dust emission. A typical minimization took 30
minutes.

\begin{figure*}
  \centering

    \includegraphics[width=0.45\textwidth]{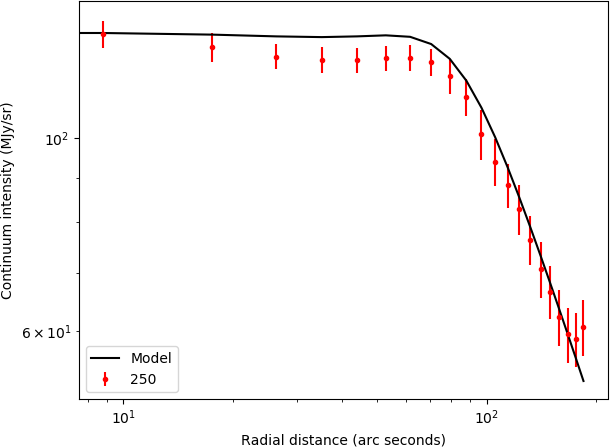}\hfill%
    \includegraphics[width=0.45\textwidth]{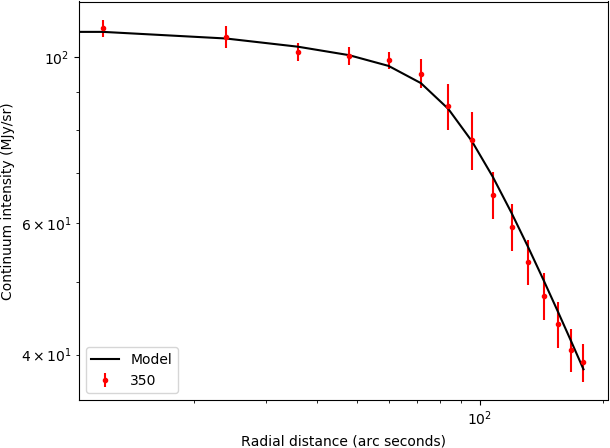}\\
   \includegraphics[width=0.45\textwidth]{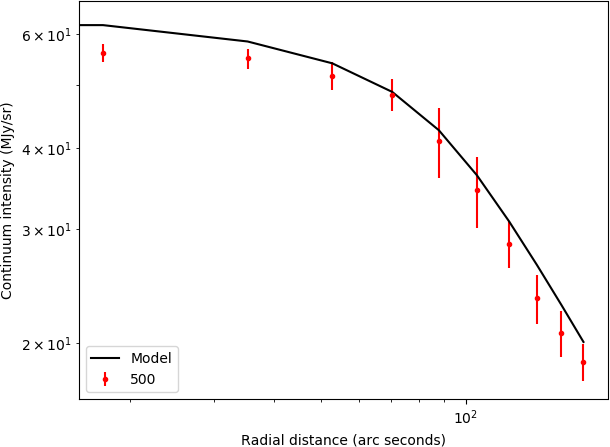}\hfill%
    \includegraphics[width=0.45\textwidth]{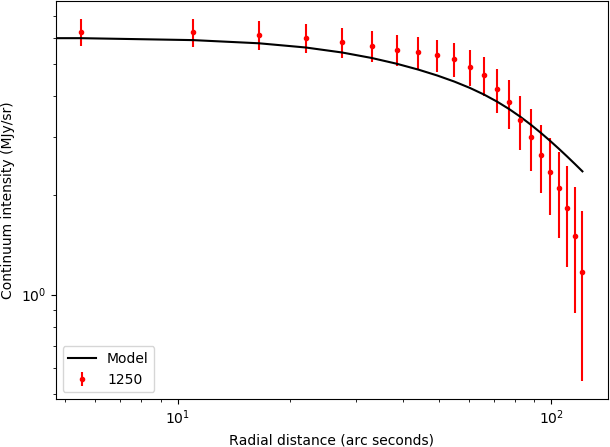}\\
      
  \caption{Continuum fit results along with the radial profiles for
    the continuum emission. The emission at 250~$\mu$m (top left), the
    emission at 350~$\mu$m (top right), the emission at 500~$\mu$m (bottom left), and the
    emission at 1.25 mm (bottom right). The black lines represent the emission
    computed from the best-fit model, while the red dots represent the
     continuum intensity along the southwest cut described
    in Section \ref{sec:cmodel}. The error bars displayed are the
    quadratic sum of the mean rms noise for each map and the standard
    deviation of all the pixels within one beam.}
  \label{fig:contmcmcfitres}
\end{figure*}

The best-fit parameters are summarized in \refdust, and the
corresponding best-fit models are compared to the observed radial
intensity profiles in Fig.~\ref{fig:contmcmcfitres}. The probability
distribution for the parameters is shown in
Fig.~\ref{fig:contmcmccorner}.

Our results may be compared with the core properties obtained in the
previous study of \cite{tafalla2004} based on a single 1.3~mm
IRAM/MAMBO map. These authors found a remarkably steep profile, with
$\alpha=3.5$. In our analysis, the main difference is precisely the
slope of the density profile in the envelope, which we find to be
significantly shallower, with $\alpha=2.2$. Nevertheless, our value is
closer to what would be expected from theoretical considerations of
spheres undergoing gravitational collapse, either isothermal or not,
which predict that the density profile in the envelope quickly
converges toward $r^{-2}$ (\citealt{larson1969} and
\citealt{lesaffre2005}). One likely explanation is that the 1.3~mm
emission essentially traces the cold dust emission from the innermost
regions of the core rather than the warmer envelope, which emits
mostly at smaller wavelengths. In addition, compared to
Herschel/SPIRE, the dual-beam observing mode used with the IRAM/MAMBO
instrument makes it less sensitive to the extended, shallow
regions. The decrease of $\alpha$ toward the theoretical value of 2 is
therefore expected from both observational and theoretical
considerations.

The other parameters of the density profile are close to those from
the analysis of \citeauthor{tafalla2004}. In particular, the inner
plateau is smaller in our study, $r_0=47\arcsec$ instead of 75\arcsec,
which can be explained by the fact we used a non-uniform dust
temperature. Attempts with a uniform temperature, although not
satisfactory, give a plateau size of 65\arcsec. A
large plateau is needed to compensate for the lower temperature in the
envelope than in our non-uniform temperature profile.

The spectral index we obtain is $\beta$=1.6. This low value is still
within the expected range of 1.6 to 2.4 for prestellar cores
\citep{makiwa2016}, but is significantly lower than the value of
$2.3\pm0.4$ measured in a more evolved core such as L1544
\citep{chacon-tanarro2017}. In a comprehensive analysis of the 1.2 and
2~mm dust emission from prestellar and protostellar cores, a broad
range of $\beta$ was obtained, showing a trend toward a decrease with
evolutionary stage \citep{bracco2017}. This may be indicative of
variations of the dust size distribution and/or composition, although
this could also be explained by variations of the mass absorption
coefficient. Interestingly, low values encompassing $\beta=1.6$ are
found in one prestellar core (Miz-2). We also note that low values
of $\beta$ could be due to dust temperatures above 10~K, such as in
L1498 based on our study, in contrast with temperatures below 10~K
found in the hundredfold denser L1544 core \citep{keto2010}.

\section{Emission lines from HCN and isotopologues}
\label{sec:lmodel}

To model the emission spectra of HCN and its isotopologues along the
southwest cut shown in Fig.~\ref{fig:cmaps}, the density profile
obtained previously was used to compute the collisional excitation
of the rotational levels of HCN and isotopologues. On the other hand,
the kinetic temperature was varied in the range 8 to 12~K and was found to
have little impact on the level populations. We therefore used a
constant kinetic temperature throughout the core and adopted a value
of 10~K \citep{tafalla2004}.  However, line radiative transfer also
involves the knowledge of the velocity field and of the line
broadening. This is particularly important in the present study of HCN
lines because of so-called anomalous hyperfine ratios.

\subsection{Radial velocity profile}
\label{sec:colvel}

Our radial velocity profile, shown in Fig.~\ref{fig:vel_sigma},
simulates that of a core at the earliest stages of collapse, in which
the outer parts move inward (\citealt{foster1993} and
\citealt{lesaffre2005}), and it reproduces the trends found in other
studies \citep{keto2010}. In our simple analytical model, the velocity
profile is given by
\begin{equation}
  \label{eq:lesaffrevcollapse}
  V(r) = V_c \exp[-(r-r_V)^2 / (2\Delta r_V^2)],
\end{equation}
with $V_c$ the maximum collapse velocity, $r_V$ the radius at which
the collapse is equal to $V_c$ , and $\Delta r_V$ the dispersion of the
collapse velocity profile. This simple parameterization is particularly
well suited to describe various stages of dense core evolution. A
detailed study of the different types of velocity profiles tested
is provided in Appendix \ref{sec:sigmavel}.

\subsection{Velocity dispersion profile}
\label{sec:nonthermal}

It is well established that emission lines become narrower toward the
inner regions of prestellar cores. This is generally interpreted as an
indication of turbulence dissipation in cores (\citealt{goodman1998}
and \citealt{falgarone1998}), from turbulence-dominated motions in the
ambient molecular cloud to almost thermally dominated line widths in
the inner parts, as observed in \object{L183} or L1506
(\citealt{falgarone1998}, \citealt{hilyblant2008} and
\citealt{pagani2010}). In the case of L1498, the FWHM of HC\fifn(1-0)
is 0.22~km/s, while the thermal contribution at 10~K is 0.13\kms,
indicating that nonthermal broadening no longer dominates. On the
other hand, the FWHM in the embedding molecular cloud in the vicinity
of L1498 is 0.63\kms\ \citep{tafalla2006}.

We therefore introduced a radial dependence of the nonthermal
broadening and adopted the following analytical formulation:
\begin{equation}
  \sigma_{\rm nth}(r) = \sigma_0 + \frac{\sigma_{\rm
      ext}-\sigma_0}{\pi}\left[\frac{\pi}{2} +
    \tanh\left(\frac{r-r_{j}}{\Delta r_{j}}\right)\right],
  \label{eq:dispersion}
\end{equation}
where $\sigma_0$ and $\sigma_{\rm ext}$ are the
nonthermal velocity dispersion at the center of the core and in the
ambient cloud, respectively. Details on the effects of the nonthermal broadening on
the line profiles are described in Appendix \ref{sec:sigmavel}.

\begin{table}[t]
  \centering
  \caption{\label{tab:cresults}Our best source model (density, dust
    properties) from the MCMC minimization of the dust emission
    maps. Note that 50\arcsec$\approx0.04$ pc at 140 pc. Quoted are
    the median values, while the uncertainties are the 16\% and 86\%
    quantiles.}
  \begin{tabular}{lcr}
    \toprule
    Parameter & Value & Unit \\
    \midrule
    $n_0$ & (1.00$\pm0.16$)\tdix{5} & \cm{-3}\\
    $r_0$ & 47$\pm$6 & arcsec\\
    $\alpha$ & 2.2$\pm$0.1 & \\
    \tempin &  9.8$\pm$0.5 & K \\
    \tempout& 14.6$\pm$0.3 & K \\
    $r_d$&   61$\pm$3 & arcsec \\
    $\Delta r_d$ & 26$_{-7}^{+10}$ & arcsec \\
    $\beta$ & 1.56$\pm$0.04 & \\
    \bottomrule
  \end{tabular}
\end{table}

\begin{table}[t]
  \centering
  \caption{Resulting parameters from our MCMC line fitting, adopting
    the source model from Table~\ref{tab:cresults}. Note that
    50\arcsec$\approx0.04$~pc at 140 pc. Quoted are the median values,
    while the uncertainties are the 16\% and 86\% quantiles.}
  \begin{tabular}{lcr}
    \toprule
    Parameter & Value & Unit \\
    \midrule
    $V_c$& -0.26$\pm$0.02 & \kms \\
    $r_V$ & 290$\pm8$ & arcsec\\
    $\Delta r_V$& $55\pm8$ & arcsec\\
    $\sigma_0$ & 0.046$\pm0.003$ & \kms\\
    $\sigma_{\mathrm{ ext}}$ & 0.25$\pm0.08$ & \kms\\
    $r_j$ & 320$\pm14$ & arcsec \\
    $\Delta r_j$ & $78\pm14$ & arcsec\\
    $r_1$ & 33$\pm$8& arcsec\\
    $X_0$ & (2.6$\pm0.2$)\tdix{-9}\\% relative to \ce{H2}\\
    $X_1$ & (6.1$\pm0.4$)\tdix{-9}\\% relative to \ce{H2}\\    
    $\eta$(\trhcn)\hspace{5ex} & 45$\pm$3\\% relative to \ce{H2}\\
    $\eta$(\fihcn) & 338$\pm$28\\% relative to \ce{H2}\\
    \bottomrule
  \end{tabular}
  \label{tab:lresults}
\end{table}

\subsection{Abundances}

Abundances in dense cores usually evidence a radial dependence, with a
clear tendency toward depletion in the inner parts, interpreted as a
signature of ice formation on grains \citep{tafalla2006}. Although
this effect is particularly dramatic for carbon-bearing species at
densities of a few \dix{4}\ccc, nitrogen-bearing species, including CN,
seem to remain in the gas phase at higher
densities \citep{hilyblant2008}. Whether HCN does follow the trend of
carbon-bearing species or not remains unclear, although observations
suggest it does not \citep{hilyblant2010}. We thus allowed for depletion
of HCN (and isotopologues) in a central region of radius $r_1$, and
assumed a stepwise abundance profile.

One fundamental assumption (although a usual one) in our model is
that HCN and its isotopologues are cospatial. Based on chemical
considerations, the formation and destruction of these species are
driven by the same bimolecular reactive collisions. We also assumed
that the abundance profiles of HCN, H\thcn, and HC\fifn\  only differ
by a multiplicative factor. Radial variation of the isotopic ratios
could result from temperature gradients driving the efficiency of
chemical fractionation. Kinetic temperature variations, from 12 to 6~K
in the center, were reported in the L1544 core
\citep{crapsi2007}. In L1498, however, no such variation were observed
\citep{tafalla2006}, probably because of the earlier stage of
condensation of this core compared to L1544. We thus consider our
assumptions at least reasonable ones.

Finally, the abundance profiles of the three isotopologues are of the form
\begin{equation}
  X(r) =
  \begin{cases}
    {X_0}/{\eta} &  r<r_1\\
    {X_1}/{\eta} &  r\geqslant r_1\\
  \end{cases}
,\end{equation}
where $X_0$ and $X_1$ are the abundances of the main isotopologue at
radii smaller and larger than $r_1$, the depletion radius,
respectively, and $\eta$ is a scaling factor, one for each of the
\thc\ and \fifn\ isotopologue.

\subsection{Modeling emerging spectra with \alico}
\label{sec:mcmc}

To model the emerging spectra of HCN and its isotopologues, we have
used the accelerated lambda iteration code \alico.  \alico\ is a
python wrapping over the \texttt{1Dart} code \citep{daniel2008}.
\alico\ handles line overlaps, including hyperfine transitions, which
is fundamental for the reproduction of the HCN spectra from cold cores
(\citealt{gottlieb1975}, \citealt{guilloteau1981} and
\citealt{gonzalez-alfonso1993}).  \alico\ takes as input collisional
coefficients, spectroscopic information, and a physical structure
assuming spherical symmetry, which consists of radial profiles of the
density, kinetic temperature, abundances, a radial velocity, and
velocity dispersion.

This work benefitted from the most accurate hyperfine HCN-\ce{H2}
collisional rates of Hernandez-Vera et al. (in preparation).  These
rates have been computed using a full quantum time-independent
scattering method.  The so-called "close-coupling" method was combined
with the potential energy surface (PES) that was computed by
\cite{denis-alpizar2013} at the explicitly correlated coupled-cluster
with a single-, double-, and perturbative triple-excitation level of
theory.  The accuracy of this PES was checked by comparing the
measured ro-vibrational spectra of the HCN-\ce{H2} complex with
bound-state calculations.  Excellent agreement was found, with
differences between observed and calculated transition frequencies
lower than 0.5\%, suggesting that the PES can be used with confidence
to compute collisional rate coefficients.  The rotational rate
coefficients are described in \cite{hernandezvera2017}, where data are
provided for the lowest 26 rotational levels and temperatures in the
range 5-500~K for both para- and ortho-\ce{H2} as colliders.  The
hyperfine collisional rates were obtained by Hernandez-Vera et al. (in
preparation) using the so-called almost exact "recoupling" method for
the lowest 25 hyperfine levels and temperatures in the range 5-30~K
for para-\ce{H2}.

For the spectral line fitting, we used the density profile derived in
Section \ref{sec:densprof}, with the addition of a constant term to
account for the presence of a low-density envelope.  This low density
envelope is seen in emission in CO spectra \citep[see App.~B
in][]{tafalla2006} and is partly responsible for the self-absorption
seen in the HCN(1-0) observations. This absorbing layer is indeed
crucial to obtain good fits to the line profiles, and especially to
reproduce the hyperfine ratios of HCN. The density profile is thus
described by:
\begin{equation}
  \nhh(r) = \frac{n_0}{1+(r/r_0)^\alpha} +n_{\mathrm{ext}}
  \label{eq:nh}
\end{equation}
where all the parameters are taken from our continuum fit
(Section~\ref{sec:cmodel}), except for $n_{\mathrm{ext}}$, which is
taken to be 500~\cm{-3}, that is, $n_{\ce{H}}$= 1000~\cm{-3}
(\citealt{hilyblant2007} and \citealt{lippok2016}). The remaining
parameters collapse velocity, nonthermal broadening, and abundances
were all left to vary within bound intervals.  All values within the
bound intervals were considered to be equally probable.  The bound
intervals along with the initial values were passed to an MCMC sampler
that indipendently searched for the minima within the bound intervals
for all parameters.  A detailed description of this fitting procedure
is given in Appendix \ref{sec:alicomcmc}.

\begin{figure*}
  %\begin{left}
  a)\\
    \includegraphics[width=\textwidth]{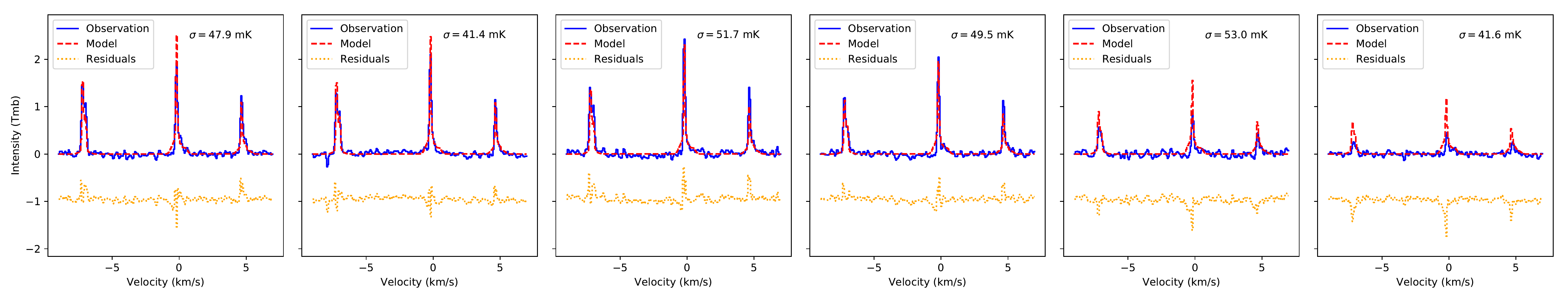}\\
  b)\\
   \includegraphics[width=\textwidth]{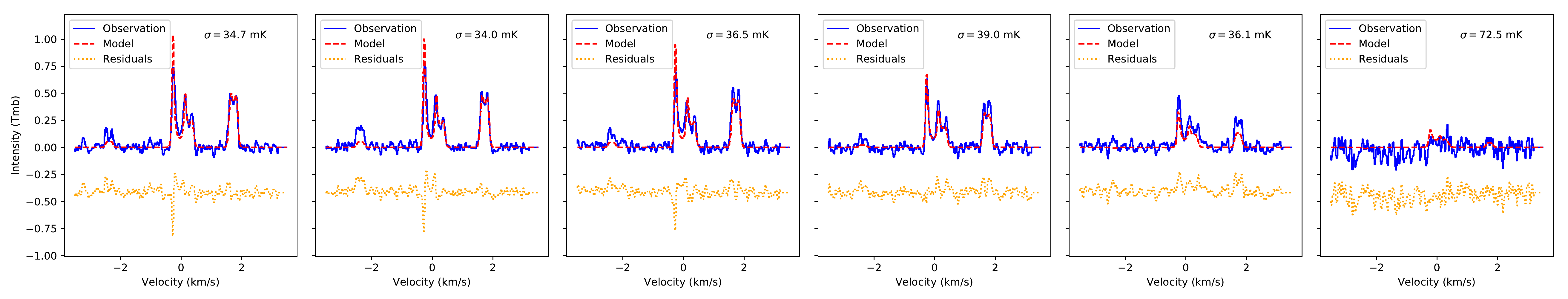}\\
  c)\\
   \includegraphics[width=\textwidth]{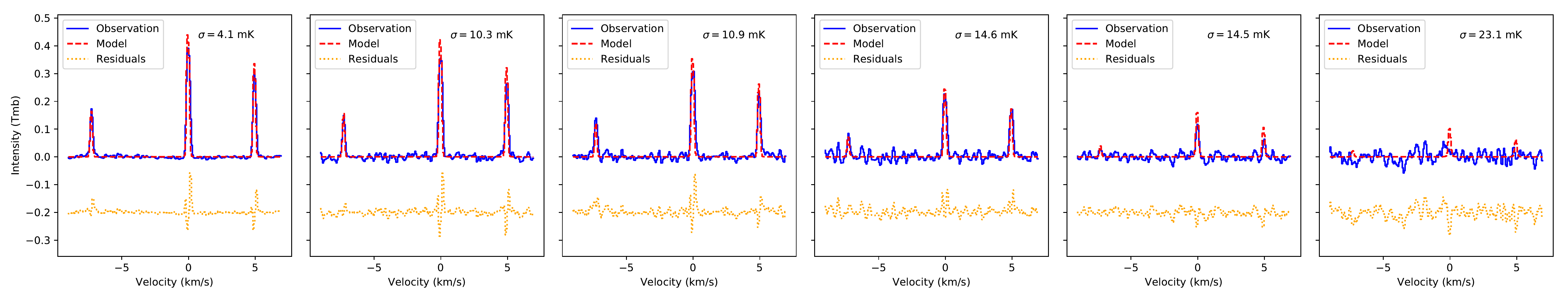}\\
  d)\\
   \includegraphics[width=\textwidth]{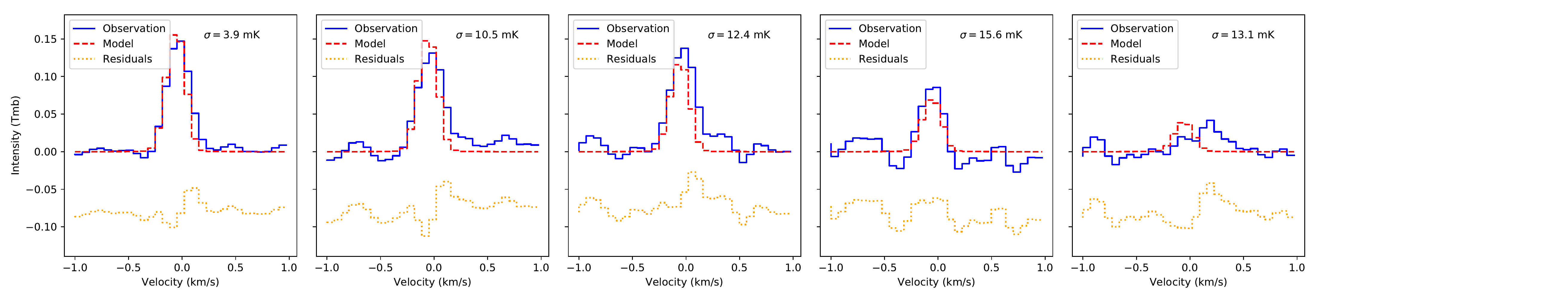}

   \caption{Best-fit \texttt{ALICO} models (red) to the HCN(1-0) (a)
     ), HCN(3-2) (b) ), \trhcn(1-0) (c) ), and \fihcn(1-0) (d) )
     spectra of L1498 (blue), along with the residuals (orange). The
     spectra are arranged in increasing distance to the continuum peak
     from the left to the right.  a): (0,0)", (-10,-20)", (-20,-40)",
     (-30,-60)", (-40,-80)", and (-50,-100)"; b): (0,0)", (-5,-10)",
     (-10,-20)", (-20,-40)", (-30,-60)", and (-40,-80)"; c): (0,0)",
     (-10,-20)", (-20,-40)", (-30,-60)", (-40,-80)", (-50,-100)"; and d):
     (0,0)", (-10,-20)", (-20,-40)", (-30,-60)", and (-40,-80)".  }
    \label{fig:alicofits}
\end{figure*}

\section{Results and discussion}
\label{sec:results}

The best-fitting values for the collapse velocity, line broadening,
and abundances are summarized in Table \ref{tab:lresults}, and the
corresponding spectra are shown in
Figure~\ref{fig:alicofits}.  

\subsection{HCN line profiles}

The line profiles of both HCN transitions are well reproduced,
including the self-absorption features and the hyperfine anomalies.
The self-absorption features are reproduced in all hyperfine
components where they show up, in both intensity and width.  The
red-blue asymmetry of the self-absorption features in HCN(1-0) is well
reproduced, along with the lack of red-blue asymmetry of the HCN(3-2)
transition.

\begin{figure}
  \begin{center}
    \includegraphics[width=\hsize]{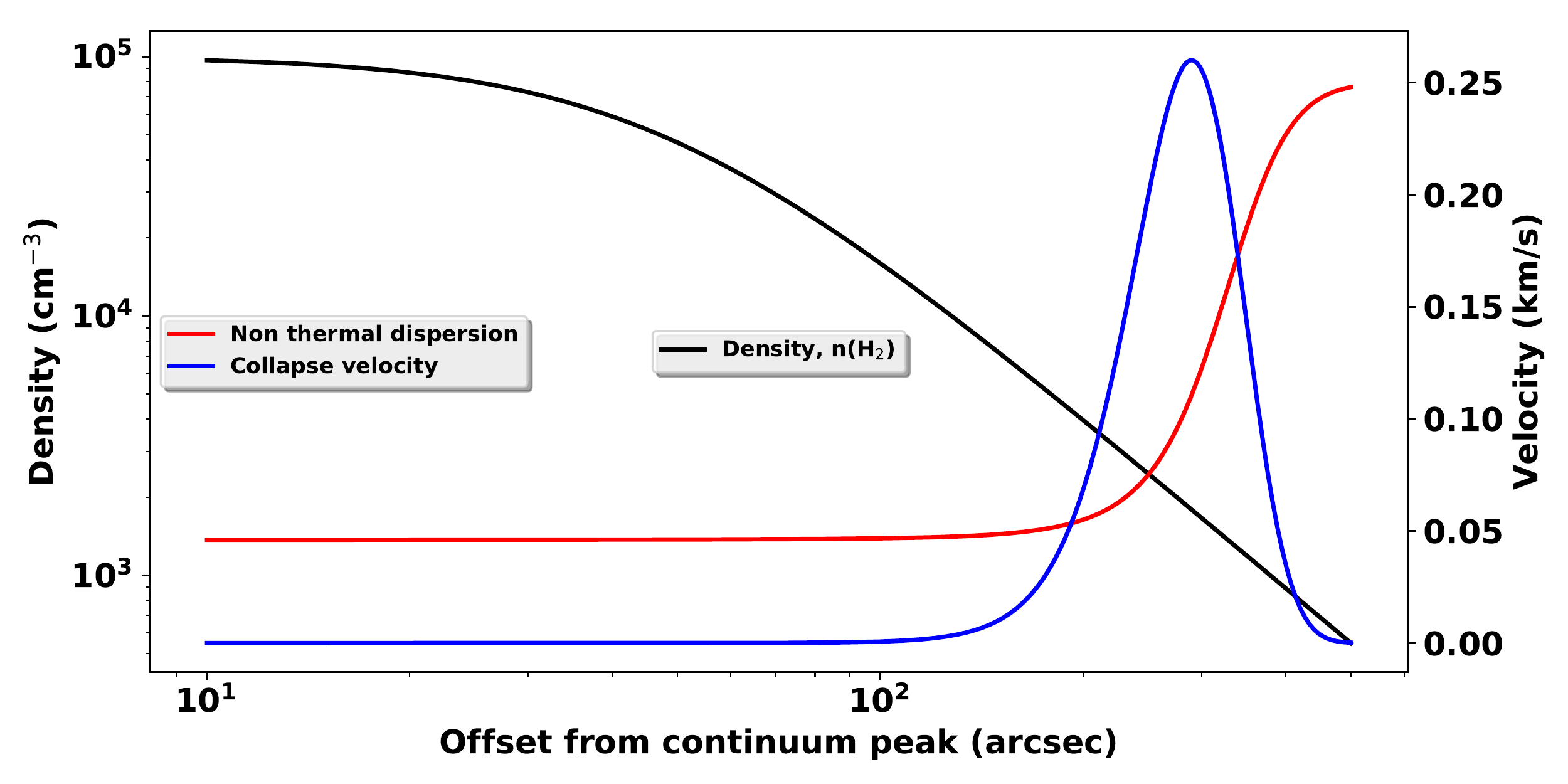}
    \caption{Collapse velocity and nonthermal velocity dispersion
      profiles derived from the fitting of the HCN line profiles.}
    \label{fig:vel_sigma}
  \end{center}
\end{figure}

The absence of collapse signatures in the HCN(3-2) spectra is found to
provide strong constraints on the collapse velocity profile, and
especially on $r_V$ and $\Delta r_V$. However, the best-fit value of
$r_V\approx300$\arcsec\ would correspond to a collapsing layer located
well within the low-density ($\nhh\approx\dix{3}$\ccc) envelope of the
core, which is unphysical. Nevertheless, an inward-moving layer is
required to reproduce the skewed profile of the HCN(1-0) emission
line. More specifically, such inward motion has to occur in a region
where the population of the \jlev{2} is small, thus putting it in the
envelope of the core, or in the surrounding molecular cloud. We note
that the peak value of the inward velocity profile, 0.26\kms, is close
to that of the foreground layer, 0.35\kms, proposed by
\cite{tafalla2006} to explain features seen in both HCN and CO spectra
of L1498. The distance of the moving parcel of our model from the
center of the cloud suggests that the red-blue asymmetries observed in the
HCN(1-0) spectra are caused by such a foreground layer and that L1498
is in fact a stationary dense core, that is to say, it does not undergo collapse. The
velocity and nonthermal velocity dispersion profiles derived from the
line profile fitting is shown in Figure~\ref{fig:vel_sigma}.

\subsection{HCN isotopic ratios}

Based on the best model, the median values of the HCN/HC\fifn,
HCN/\trhcn, and \trhcn/HC\fifn\ abundance ratios are 338$\pm$28,
45$\pm$3, and 7.5$\pm$0.8, respectively, where the uncertainties are
the 16\% and 84\% quantiles. The accuracy is thus on the order of 10\%
for the three ratios, indicating that our results have reached, and do
not go beyond, the telescope calibration accuracy.

The nitrogen isotopic ratio in HCN is thus found in remarkable
agreement with the elemental ratio of 330$\pm$30 in the local ISM.
This implies that HCN is not fractionated in nitrogen in L1498. If
this result is confirmed toward other prestellar cores, the
fractionation of HCN seen in protoplanetary disks must therefore occur
at a later stage of the star formation sequence, unless isotopic
exchanges in ices take place efficiently during the core
contraction. On the other hand, the HCN is significantly enriched in
\thc \ compared to the elemental carbon isotopic ratio of \cratio =
68$\pm$15 \citep{milam2005}. This \thc\ enrichment is similar to the
one reported in the more evolved Barnard B1 core (\citealt{daniel2013}
and \citealt{gerin2015}), where a ratio of 30$\pm$7 in HCN was
measured. Regarding \trhcn\ and HC\fifn, our derived ratio is
consistent with \citet{ikeda2002}, who obtained
\trhcn/\fihcn=7.9$\pm$0.9, assuming a single excitation
temperature. The H\thcn/HC\fifn\ ratio is only marginally consistent
with the value of 5.5$\pm$1.0 in B1.

Assuming that HCN/H\thcn=45 is representative of dense, shielded gas,
this translates into a $\sim35\%$ decrease of the \nratio\ in HCN
obtained through the double-isotopic method assuming the elemental
ratio of 68. In the sample of five protoplanetary disks studied in
\citet{guzman2017}, the average HCN/\fihcn\ ratio of 111$\pm$19 thus
would become 71$\pm$13. In the L1544 and L183 dense cores, the
HCN/HC\fifn\ ratio becomes 157$\pm$37 and 131$\pm$29, respectively.
These new values are in harmony with the directly measured HCN/\fihcn\
ratio in the B1 cloud, 165$\pm$30 \citep{daniel2013}, although
the value was discarded by the authors considering that the column
density of HCN could not be reliably estimated. These revised
HCN/\fihcn\ ratios in L1544 and L183, which are a factor 2--2.5 lower
than the elemental ratio, would thus be the signatures of a secondary,
\fifn-rich reservoir of nitrogen at the prestellar core stage. From a
cosmochemical perspective, such \fifn-enrichments are consistent with,
although slightly below, what would be expected for comets forming in
disks within these cores (HB17). In addition, the revised average
ratio in disks would also be close to the largest enrichments measured
in hotspots within solar system chondrites \citep{bonal2010}.

The new ratios in L1544 and L183 are thus lower by the same factor
2--2.5 as the ratio in L1498 obtained in the present
study. This either indicates that there is a variability of the carbon
and/or nitrogen isotopic ratios from source to source, even when
embedded in the same large-scale environment as L1498 and L1544, or
that the carbon and/or nitrogen isotopic ratios are incorrect in L1544
and L183. Distinguishing between the two possibilities requires
detailed studies such as the one performed here in L1498.

Finally, we note that the \nratio\ in HCN in L1498 is found to agree
well with some models of chemical fractionation in dense cores
(\citealt{terzieva2000} and \citealt{roueff2015}) but disagrees with
others \citep{hilyblant2013c15n}. In particular, the
\citet{roueff2015} model, which also includes fractionation of carbon,
predicts strong depletion of HCN in \thc by factors from 1.3 to 2.4,
which is at variance with our present results. Naturally, our
\trhcn/\fihcn\ also disagrees with these models, which predict values
in the range 2.6 to 4.3, whereas we obtain 7.5$\pm$0.8. It is
therefore unclear which of the two ratios, HCN/\trhcn\ or
\trhcn/\fihcn, is actually not reproduced by these models.
Table~\ref{tab:obsvsmodels} displays a comparison of the observed HCN
isotopic ratios and the predictions from chemical models.

\begin{table*}
  \centering
  \caption{Isotopic ratios in carbon and nitrogen obtained in
    prestellar cores are compared to model predictions.}
  \label{tab:obsvsmodels}
  \begin{tabular}{lllccccr}
    \toprule
    Ratio & Source & Observed  & \multicolumn{4}{c}{Models$^\S$} & Comments \\
            &&value\dag&(1) & (2) & (3) & (4) \\
    \midrule
HCN/\trhcn & L1498 & 45$\pm$3 & - & - & - & 114 - 168 & This work \\
 & B1 & 30$_{-4}^{+7}$ &  &  &  &  &  (5) \\
HCN/\fihcn & L1544 & 140 - 360\myddag & 254 & 37 - 195 & 202 & 334 - 340 &  (6) \\
 & L183 & 140 - 250\myddag &  &  &  &  &  (6) \\
 & L1498 & 338\mypm28 &  &  &  &  & This work \\
 & B1 & 165\euplow{30}{25} &  &  &  &  &  (5) \\
\trhcn/\fihcn & L1544 & 2.0-4.5 & -  & - & - & 1.9 - 3.0 &  (6) \\
 & L183 & 2.0-3.7 &  &  &  &  &  (6) \\
 & L1498 & 7.5\mypm0.8 &  &  &  &  & This work \\
 & B1 & 5.5\mypm0.8 &  &  &  &  &  (5) \\
CN/\thc N & B1 & 50\euplow{19}{11} & - & - & - & 63 - 84 &  (5) \\
CN/C\fifn & L1544 & 510\mypm70\myddag & 252 & - & 409 & 334 - 337 &  (3) \\
 & L1498 & 476\mypm70\myddag  &  &  &  &  &  (3) \\
 & B1 & 240\euplow{135}{65} &  &  &  &  &  (5) \\
\thc N/C\fifn & L1544 & 7.5\mypm1.0 & -  & - & - & 4.0 - 5.3 &  (3) \\
 & L1498 & 7.0\mypm1.0 &  &  &  &  &  (3) \\
 & B1 & 4.8\mypm3.3 &  &  &  &  &  (5) \\
HNC/HN\thc & B1 & 20\euplow{5}{4} & - & - & - & 121 - 180 &  (5) \\
HNC/H\fifn C & B1 & 75\euplow{25}{15} & 237 & - & - & 332 - 335 &  (5) \\
HN\thc/H\fifn C & B1 & 3.8\mypm1.6 & - & - & - & 1.3 - 2.8 &  (5) \\
    \bottomrule
  \end{tabular}
  \begin{list}{}{}

  \item $^\S$ Steady-state isotopic ratios from the following models:
    (1) \cite{terzieva2000}; (2) \cite{wirstrom2012}; (3)
    \cite{hilyblant2013c15n} and (4) \cite{roueff2015}. The nitrogen
    isotopic ratios have been rescaled according to the elemental
    value of the local ISM, \nratio\ = 330\mypm30.\\
  \item \dag Observed values from (3) \cite{hilyblant2013c15n}; (5)
    \cite{daniel2013}; (6) \cite{hilyblant2013icarus} and this work.
  \item \myddag Values computed assuming \twc/\thc\ = 68.
  \end{list}
\end{table*}

\subsection{Depletion of HCN from the gas phase}

Table~\ref{tab:lresults} shows that our analysis indicates that HCN
is at most only moderately depleted in the inner region of L1498.
This is at variance with the result of \citet{padovani2011}, who
suggested that HCN in L1498 is depleted by several orders of magnitude
inside a hole of radius 8\tdix{16}~cm.  In contrast, our analysis
shows that the abundance of HCN in the inner region of radius
$\sim$7\tdix{16}~cm is only a factor 2.3 lower than in the outer
parts. This small factor is likely an upper limit to the
depletion, as there could be some effect of the convolution of the
non-spherical core, especially the northeast, sharp region, by the
28\arcsec\ beam (or 6\tdix{16}~cm). More specifically, the sharp drop
in \ce{H2} column density toward the northwest may be the reason for
the decrease in the HCN average column density within the beam when
pointing toward the continuum peak. Nevertheless, a detailed analysis
of the depletion of HCN would require a non-spherical model different
from the
one used in this work.

Other studies have also reported no evidence of HCN
depletion. \citet{sohn2004} found that the intensity of the \ftr{0}{1}
hyperfine component of HCN(1-0) correlates strongly with the intensity
of \ce{N2H+}, a molecule that does not show depletion in prestellar
cores at the scales sampled by single-dish telescopes. In a detailed
analysis of the HCN emission toward two contracting cores,
\citet{lee2007} also obtained a flat HCN abundance profile in
\object{L694-2}, at 7\tdix{-9} relative to \ce{H2}, while they derived
an increase from 1.7 to 3.5 \tdix{-9} in L1197 when moving
inward. Even in the more evolved \object{L1544} core, where the
central density is 100 times higher than in L1498, \trhcn(1-0)
emission shows no hint of depletion (\citealt{hilyblant2010} and
\citealt{spezzano2017}). The result of our detailed analysis of HCN in
L1498 therefore shows that unlike previous claims, this core behaves
like other cores, and also that HCN behaves like CN and HNC
\citep{hilyblant2008, hilyblant2010}.

\subsection{Accurate measurements of isotopic ratios}

As a basis of comparison, we have also computed the column densities of
HCN and its isotopologues by fitting the spectra to spectrally
resolved escape probability models using the code \texttt{RADEX}
\citep{vandertak2007} and the same MCMC technique as for the fits with
\alico.  The resulting column densities for HCN and its isotopologues
are N(HCN) = 1.1\mypm0.3\tdix{14}\cm{-2}; N(\trhcn) =
3.2\mypm1.1\tdix{12}\cm{-2} , and N(\fihcn) =
3.9\mypm1.6\tdix{11}\cm{-2}.  These column densities translate into
the abundance ratios HCN/\trhcn\ = 36\mypm7; HCN/\fihcn\ = 295\mypm75,
and \trhcn/\fihcn\ = 8.2\mypm2.4.

The \trhcn/\fihcn\ ratio derived from the escape probability fit and
the complete radiative transfer models are consistent, 8.2$\pm$2.4 and
7.5$\pm$0.8, respectively.  The main difference between the two
methods in this case are the uncertainties.  This indicates that
escape-probability-derived \trhcn/\fihcn\ ratios are reliable.  For
the case of the main isotopologue, the picture is less promising for
escape probability methods.  The inability to treat hyperfine overlaps
at the excitation level precludes them from being capable of
accurately reproducing the HCN line profiles.  Therefore the
derivation of the column density of the main isotopologue through this
method should be regarded with caution.  To completely reproduce the
HCN line profiles and consequently derive an accurate HCN isotopic
ratio, a complete radiative transfer model is necessary.  The main
difficulty resides in the fact that most parameters for these models
are degenerate and nonlinear \citep{keto2004}.  In such a case,
$\chi^2$ statistics fails to properly measure the uncertainties of the
parameters.  To efficiently explore this parameter space, we used the
affine invariant Markov chain Monte Carlo ensemble sampler
(\citealt{goodman2010} and \citealt{emcee}).  From this, we obtained
non-arbitrary uncertainties based on the analysis of the Markov
chains.

The two main sources of uncertainties that affect measurements of
isotopic ratios are the calibration uncertainties (5-10\%) and the
uncertainties on the collisional excitation rates
\citep[$\sim$20\%,][]{stoecklin2017}. Clearly, assumptions on the
source geometry in the treatment of the radiative transfer carry
additional, systematic uncertainties that are very difficult to
quantify, however, but that may dominate in certain circumstances. The
impact of the uncertainties on the collisional excitation rates is
still not understood and needs to be studied further. Calibration
uncertainties may be mitigated if the various isotopologues can be
observed simultaneously, depending on the frontend/backend
capabilities of the observing facility. Simultaneous observations
indeed cancel out the multiplicative fluctuations, such as receiver
gains, but generally do not cancel out differential bandpass
effects. It therefore seems hypothetical to go beyond
calibration-limited accuracy on isotopic ratios. In particular, in
their analysis of \ce{HC3N} and its \thc-isotopologues,
\citet{araki2016} claimed a 1--2\% accuracy on abundance ratios based
on the spectrally resolved hyperfine structure of the $J=5-4$
rotational transition. More importantly, the single excitation
temperature assumption applied to the main and \thc\ isotopologues is
most likely the largest source of uncertainty.

\section{Conclusions}
\label{sec:conclusion}

The HCN/\fihcn\ ratio in L1498 is 338$\pm$28, which is consistent with
the bulk of nitrogen in the local ISM and chemical fractionation
models.  The HCN/\trhcn\ ratio of 45$\pm$3 is neither consistent with
the bulk of carbon in the local ISM nor with recent chemical
fractionation models. The \trhcn/\fihcn\ ratio of 7.5$\pm$0.8 is also
not consistent with measurements in other clouds, but consistent with
previous measurements in L1498. The variations in the \trhcn/\fihcn\
ratio seen in different sources indicate that efficient
fractionation of either carbon or nitrogen in HCN takes place in
prestellar cores.  These results suggest that further work is needed
in order to understand the origins of these variations on both
observational and theoretical grounds.  It is also important to note that
we did not find any evidence that HCN suffers significant gas-phase
depletion in L1498.

Regarding the physical structure of L1498, our analysis of the dust
emission is consistent with a power-law radial density profile, with
an exponent ($\alpha=2.2$) in good agreement with theoretical
expectations. Incidentally, our derived exponent indicates that the
core is collapsing, thus building-up an envelope with an exponent
$\approx2,$ although the HCN(3-2) line shows that the collapse did not
reach the innermost regions. Nevertheless, our analysis is not able to
probe collapse velocities below $\sim0.1$\kms\ such as those obtained
in hydrodynamical simulations at early times \citep[e.g.,][]{lesaffre2005}.

We also found evidence for a radial increase of the dust temperature
from $\approx10$ to 15~K, while we could not find evidence for a
correlative increase of the gas kinetic temperature. The inner
dust temperature converges toward the 10~K gas kinetic temperature,
which could indicate a better collisional coupling between the gas and
the dust in the higher density parts of the core.

The MCMC method or similar minimization methods
offer efficient means to properly explore the parameter space and
therefore derive non-arbitrary uncertainties.  This kind of detailed
mimization is capable of producing very reliable and accurate
measurements.  However, the large amount of computational time required
to fit spectra using \alico\ limits its usage. Fortunately, escape
probability and single excitation temperature methods are reasonable
assumptions for the \trhcn/\fihcn\ ratio. In any case, evidence of
deviation from the carbon or nitrogen elemental isotopic ratios can
only be obtained with more time-consuming assumptions such as the one
used in the present work.

We also emphasize that the HCN(1-0) and HCN(3-2) hyperfine anomalies
are quantitatively reproduced, which was made possible by the
availability of precise HCN-\ce{H2} collisional rates and the ability
of \alico to treat the hyperfine overlap, confirming the models for
HCN hyperfine anomalies of \cite{guilloteau1981} and
\cite{gonzalez-alfonso1993}.  Although the carbon fractionation in HCN
may not be well understood, chemical models emphasize the caveats of
indirect measurements of isotopic ratios. Recent works have suggested
that doubly substituted \ce{N2H+} could be a potential direct probe of
the nitrogen isotopic ratio \citep{dore2017}. As demonstrated in the
present work, observational, theoretical, and modeling efforts must be
combined to enable direct, reliable estimates of isotopic ratios. Such
measurements must be performed in several species (HNC, CN, etc.) in
cores and disks, and are required to distinguish between a PSN or
interstellar origin of the fractionated reservoir observed in comets.

\begin{acknowledgements}
  We thank Marco Padovani for providing us with unpublished IRAM30m
  spectra (program 031-11). We warmly thank Fabien Daniel for
  providing us with a version of his code. VM is supported by a PhD
  grant from the Brazilian space agency (Ag\^{e}ncia Espacial
  Brasileira) through the science without borders program.  The present work 
  was realized with the support of CNPQ, 
  Conselho Nacional de Desenvolvimento Científico e Tecnológico, Brazil.
  PHB  acknowledges the \emph{Institut Universitaire de France} for
  financial support. This research has made use of data from the
  Herschel Gould Belt survey (HGBS) project
  (http://gouldbelt-herschel.cea.fr [gouldbelt-herschel.cea.fr]). The
  HGBS is a Herschel Key Program jointly carried out by SPIRE
  Specialist Astronomy Group 3 (SAG 3), scientists of several
  institutes in the PACS Consortium (CEA Saclay, INAF-IFSI Rome and
  INAF-Arcetri, KU Leuven, MPIA Heidelberg), and scientists of the
  Herschel Science Center (HSC).
\end{acknowledgements}

\bibliographystyle{aa}
\bibliography{mybib}

\clearpage
\newpage
\appendix

\section{Dust emission fitting}

\begin{figure*}
  \includegraphics[width=\textwidth]{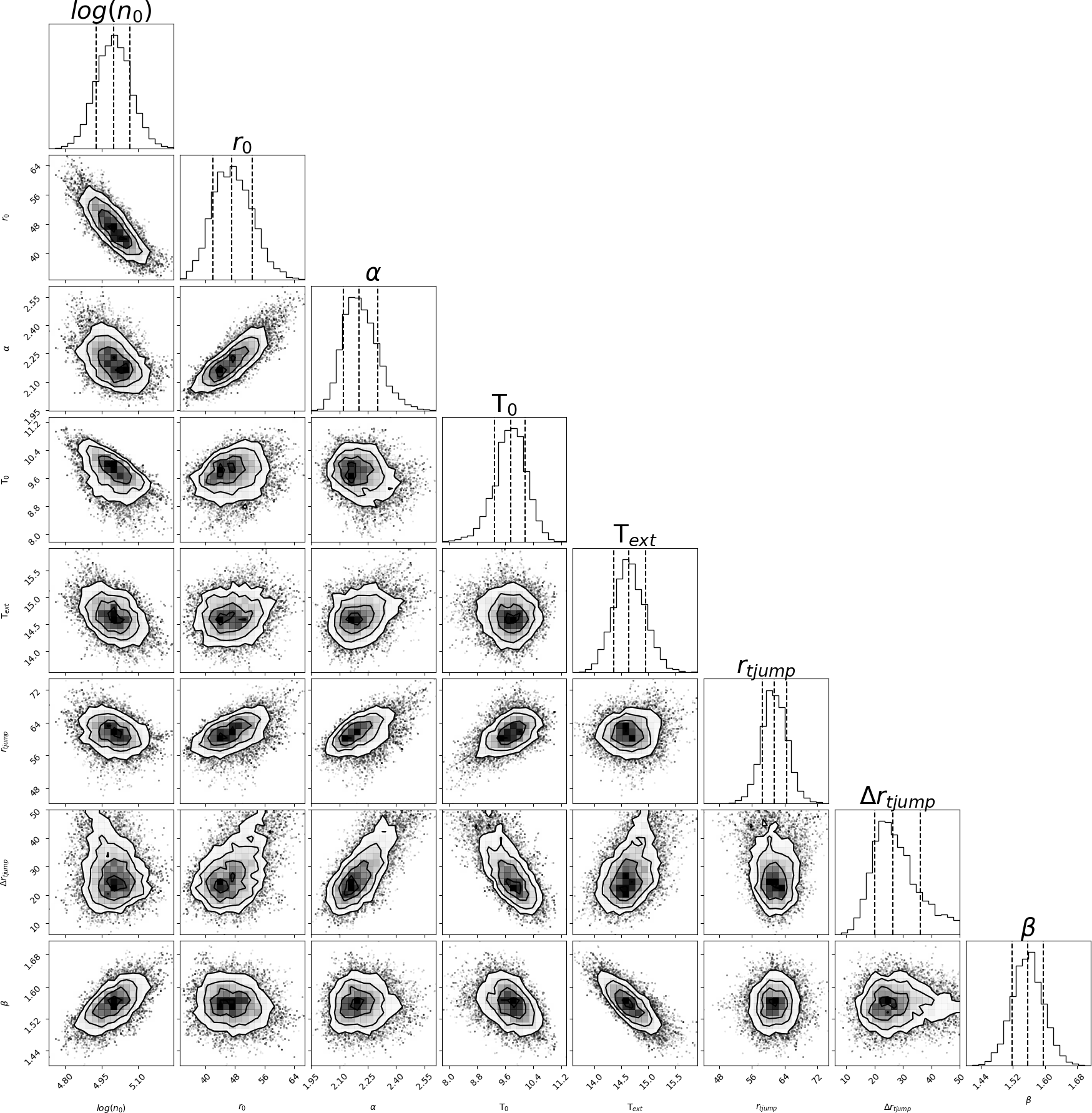}
  \caption{Probability density distribution for the parameters in the
    continuum MCMC fit, along with the scatter plots between each
    possible pair of probability density distributions. It is clearly
    visible from the scatter plots that several parameters are
    correlated.}
  \label{fig:contmcmccorner}
\end{figure*}

\subsection{Correlations between the fitting parameters}

Figure \ref{fig:contmcmccorner} shows
the correlations and anticorrelations between some of the
fit parameters.  The loose anticorrelation between $n_0$ and $r_0$
can be understood from the fact that an increase in any of them
increases the dust column density and thus the emission, so that
as one
parameter increases, the other has to decrease to preserve the observed
emission.  The positive correlation of $n_{\mathrm{ ext}}$ and
$\alpha$ comes from the fact that there is a need for some minimum
emission at larger radii, so that as $\alpha$ increases, decreasing the
line-of-sight column density at larger radii, $n_{\mathrm{ ext}}$ has
to increase to preserve it.  Not surprisingly, the tight
anticorrelation of $n_{\mathrm{ ext}}$ and $T_{\mathrm{ ext}}$ comes
from this same necessity: as $T_{\mathrm{ ext}}$ is decreased,
$n_{\mathrm{ ext}}$ has to be increased to maintain the emission, and
vice versa.  On the other hand, the positive correlation between $r_0$
and $\alpha$ is related to the slope of the emission profile decay.
Higher values of $r_0$ have to be coupled with higher values of
$\alpha$ to preserve the slope seen in the radial profiles.

The anticorrelation between $T_0$ and $T_{\mathrm{ ext}}$ can be
explained from the fact that the emission increases with both
temperatures through the term $B_{\nu}(T_{\mathrm{dust}}(r))$ in
equation \ref{eq:inudust}, therefore if $T_0$ is increased,
$T_{\mathrm{ ext}}$ has to be decreased to conserve the flux.  The
anticorrelation between $n_0$ and $T_0$ also comes from the fact that
the emission increases with $T_0$ while also being proportional to
$n_0$ through equation \ref{eq:densprof}, thus if $T_0$ is increased,
$n_0$ has to be decreased to conserve flux.  The constraints leading to
the correlation between $r_0$ and $T_{\mathrm{ ext}}$ come from the
emission at large offsets from the source, where it is dominated by
$n_{\mathrm{ ext}}$ and $T_{\mathrm{ ext}}$, but also having a
contribution from the tail of the term dependent on $r_0$ in equation
\ref{eq:densprof}, so that if flux is to be conserved, an increase in $r_0$
has to be offset by a decrease in $T_{\mathrm{ ext}}$.  The
correlations between $T_0$ and $r_0$ and $T_0$ and $n_{\mathrm{ ext}}$
arise from the anticorrelation between $T_0$ , and $T_{\mathrm{ ext}}$
acts upon $T_{\mathrm{ ext}}$ , which as a result of its anticorrelations with
$n_{\mathrm{ ext}}$ and $r_0$ , in turn cause $r_0$ and
$n_{\mathrm{ ext}}$ to be correlated with $T_0$.  The loose positive
correlation of $n_{\mathrm{ ext}}$ and $r_0$ may arise from the fact
that both $n_{\mathrm{ ext}}$ and $r_0$ present correlations with
$T_0$ and anticorrelations with $T_{\mathrm{ ext}}$, thus they are
forced to vary together by the strength of these correlations.

The anticorrelations between $\beta$ and the temperatures ($T_0$ and
$T_{\mathrm{ ext}}$) come from the fact that $\beta$ has an impact on
the relative fluxes between the different wavelengths, with high
values of beta favoring short wavelengths, which is the opposite
effect of the temperature.  To maintain the same ratio between
the fluxes at different wavelengths, a decrease in $\beta$ therefore
has to be
offset by an increase in either $T_0$ or $T_{\mathrm{ ext}}$.  The
remaining correlation, that between $\beta$ and the density
parameters ($n_0$, $n_{\mathrm{ ext}}$ and $r_0$), arises because an increase in column density requires a decrease in the dust
temperature to maintain the same emission, and with a
decrease in dust temperature, $\beta$ therefore has to increase to maintain the
flux ratio between the different continuum wavelengths.

\section{Kinematics and Nonthermal velocity dispersion in L1498}
In the following sections we use the best-fit model as the reference
model and vary only the parameters relative to the velocity profile
and nonthermal broadening profile.
\label{sec:sigmavel}
\subsection{Velocity field}

In Figure \ref{fig:hcnspecvelnovel} we show examples of spectra produced by \texttt{ALICO} with different velocity profiles and the effect of these different profiles on the HCN(1-0) HCN(3-2) line profiles.

\def\ww{0.33\hsize}
\begin{figure*} 
  \includegraphics[width=\ww]{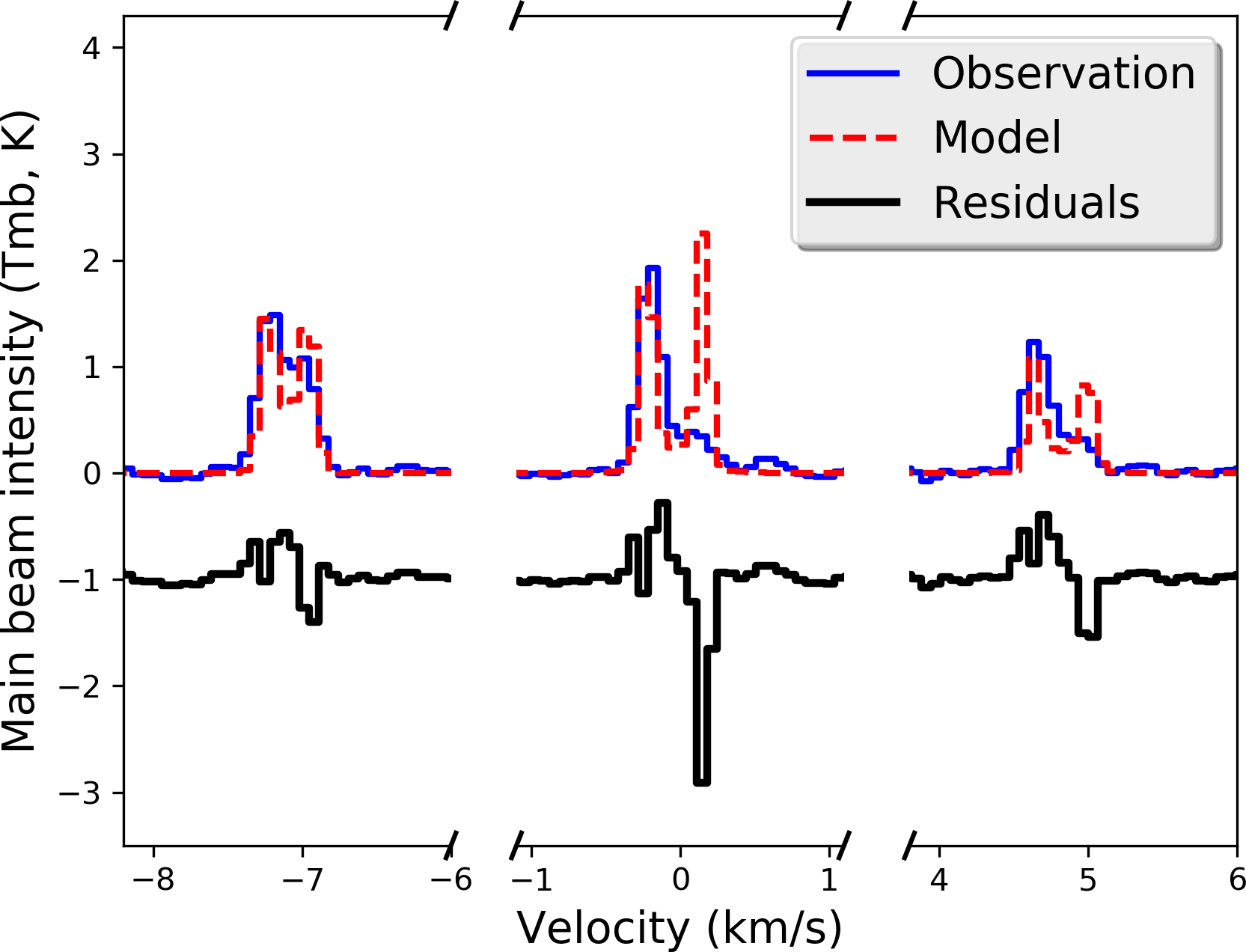}\hfill%
  \includegraphics[width=\ww]{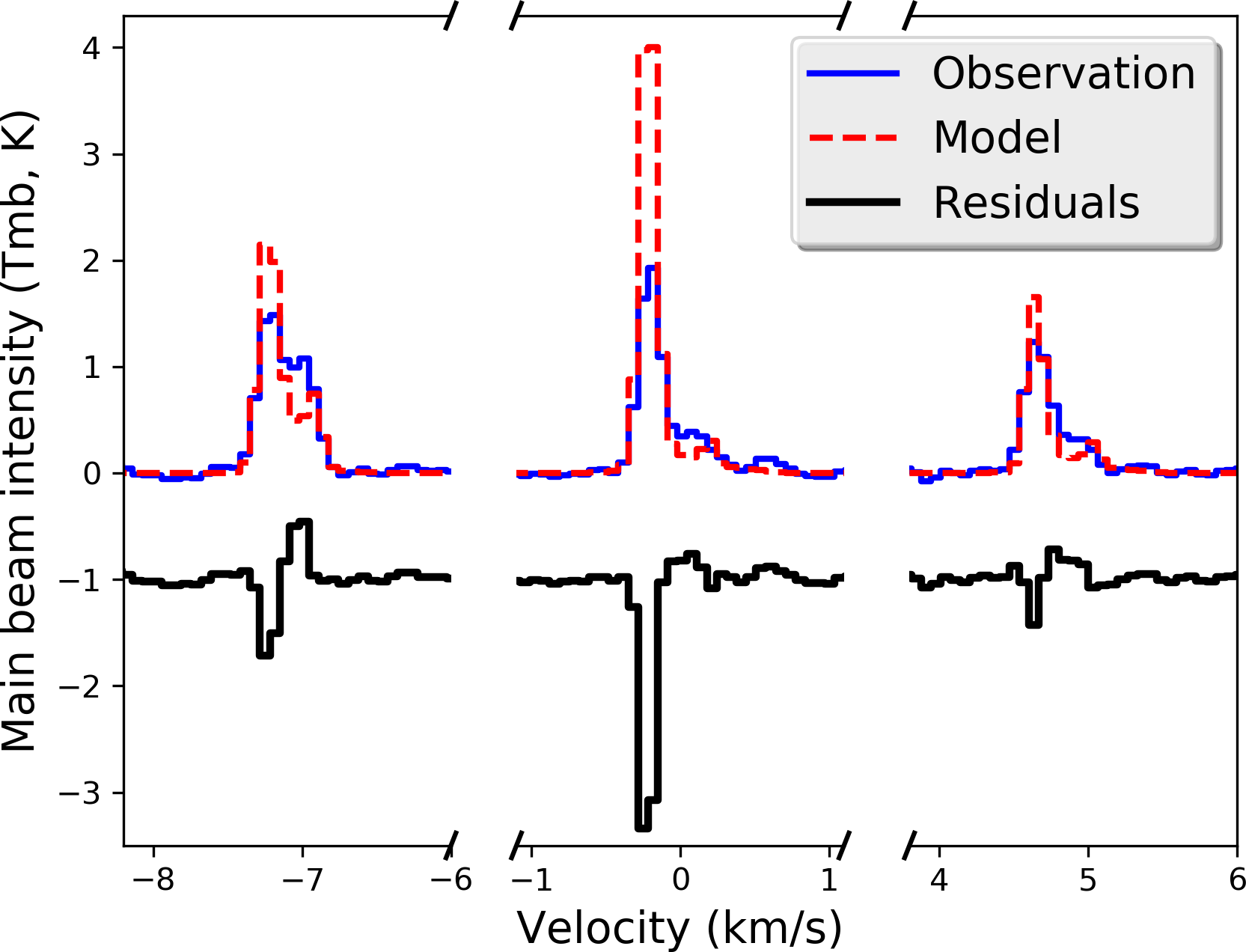}\hfill%
  \includegraphics[width=\ww]{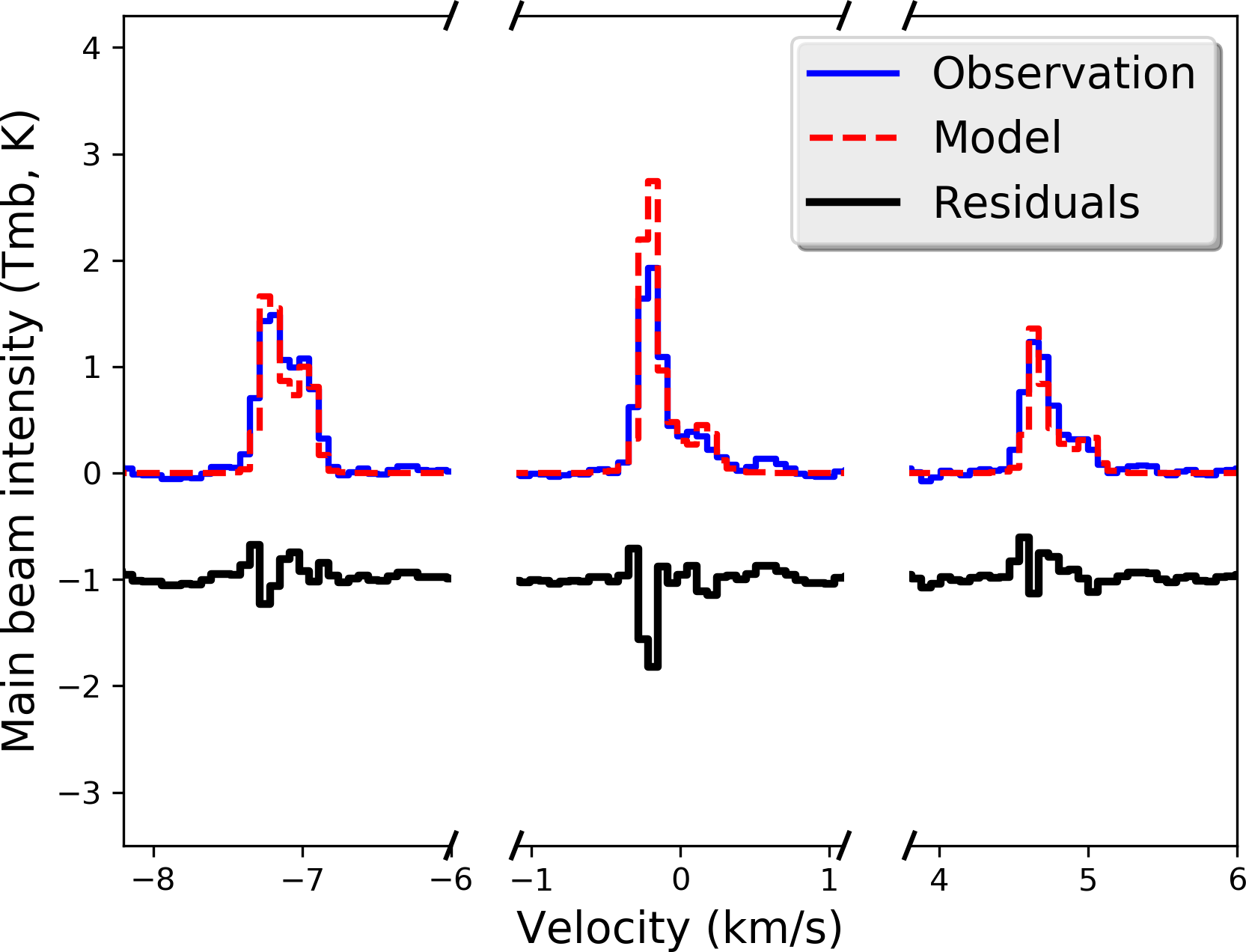}\\
  \includegraphics[width=\ww]{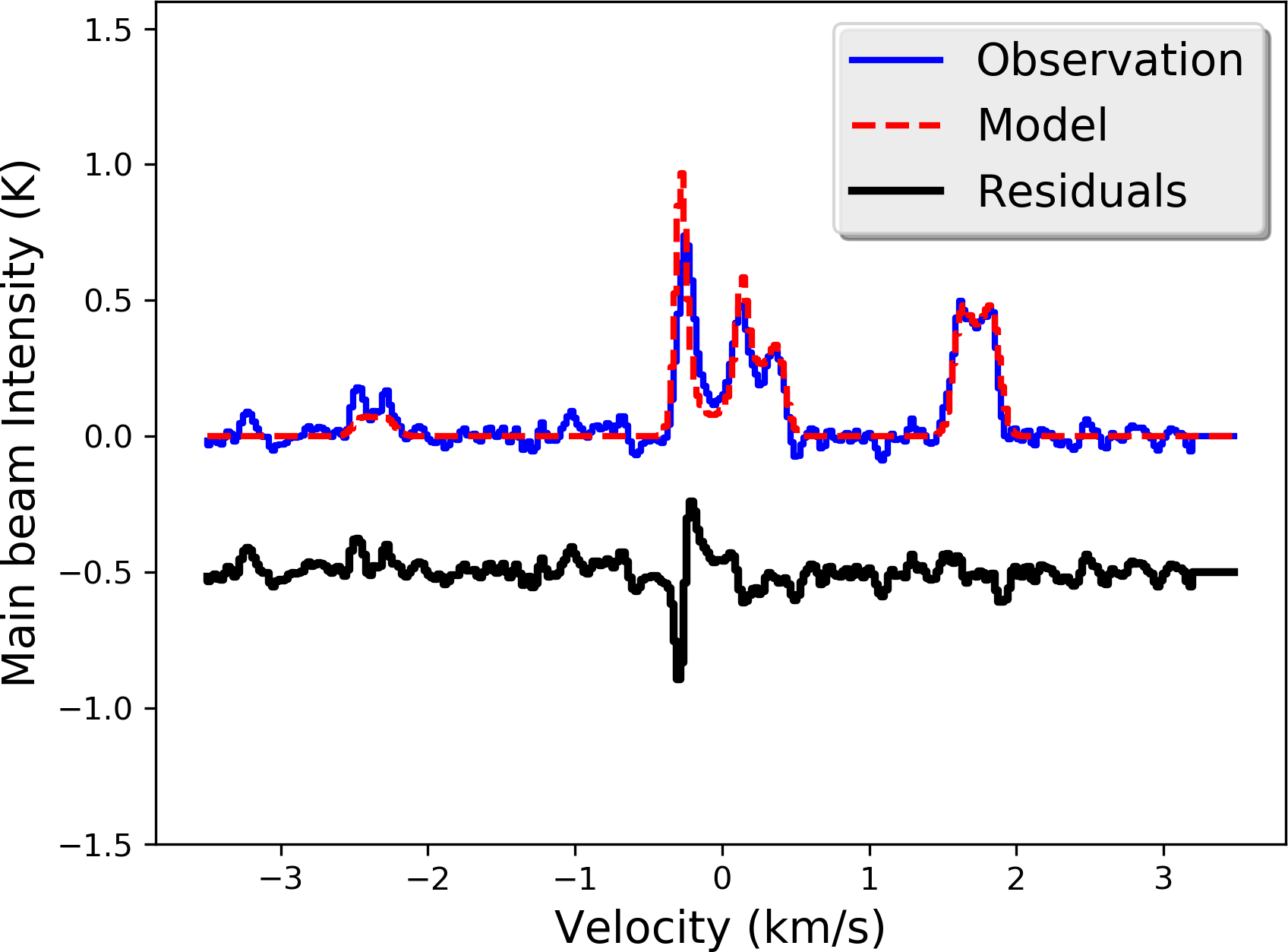}\hfill%
  \includegraphics[width=\ww]{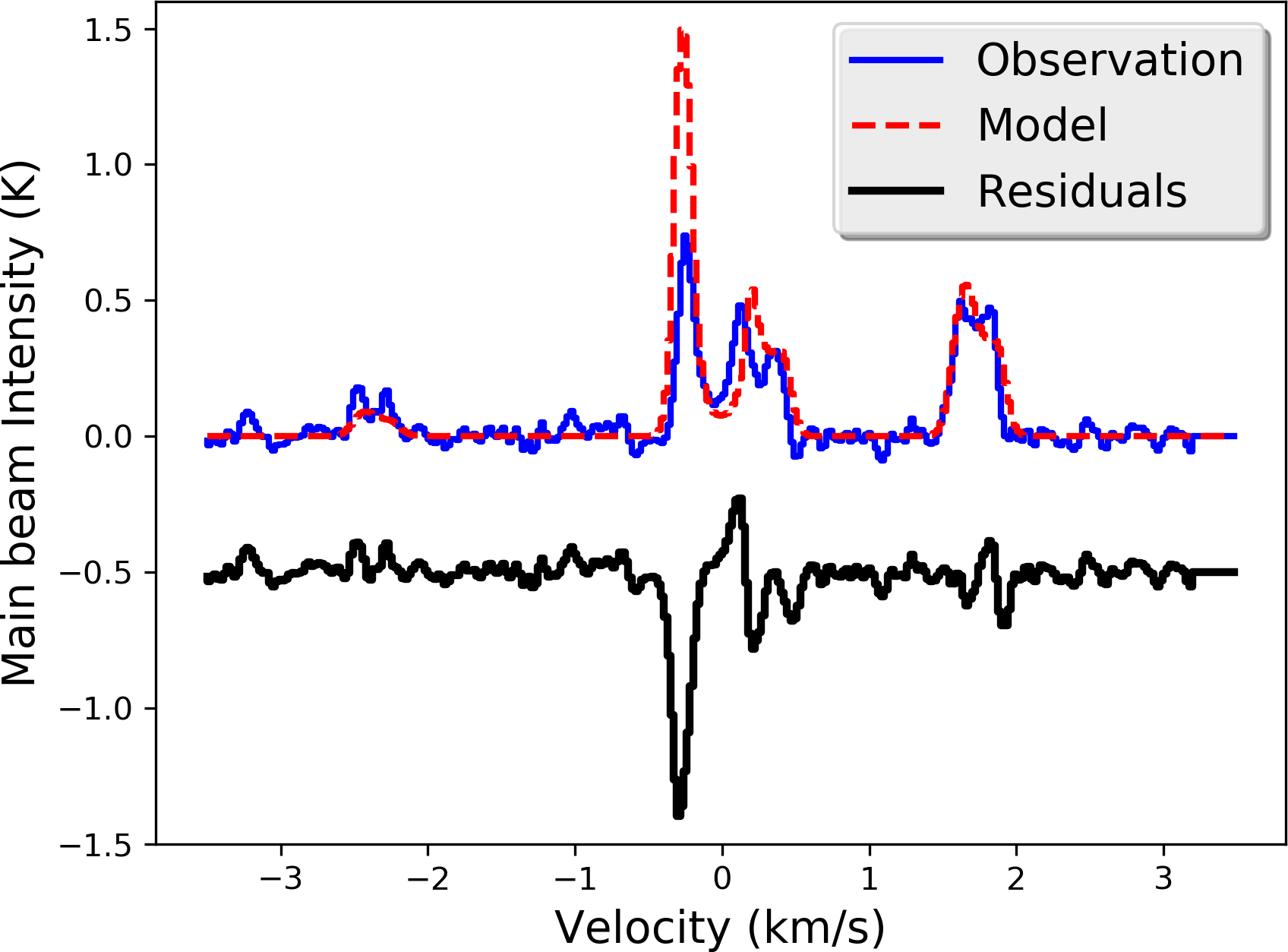}\hfill%
  \includegraphics[width=\ww]{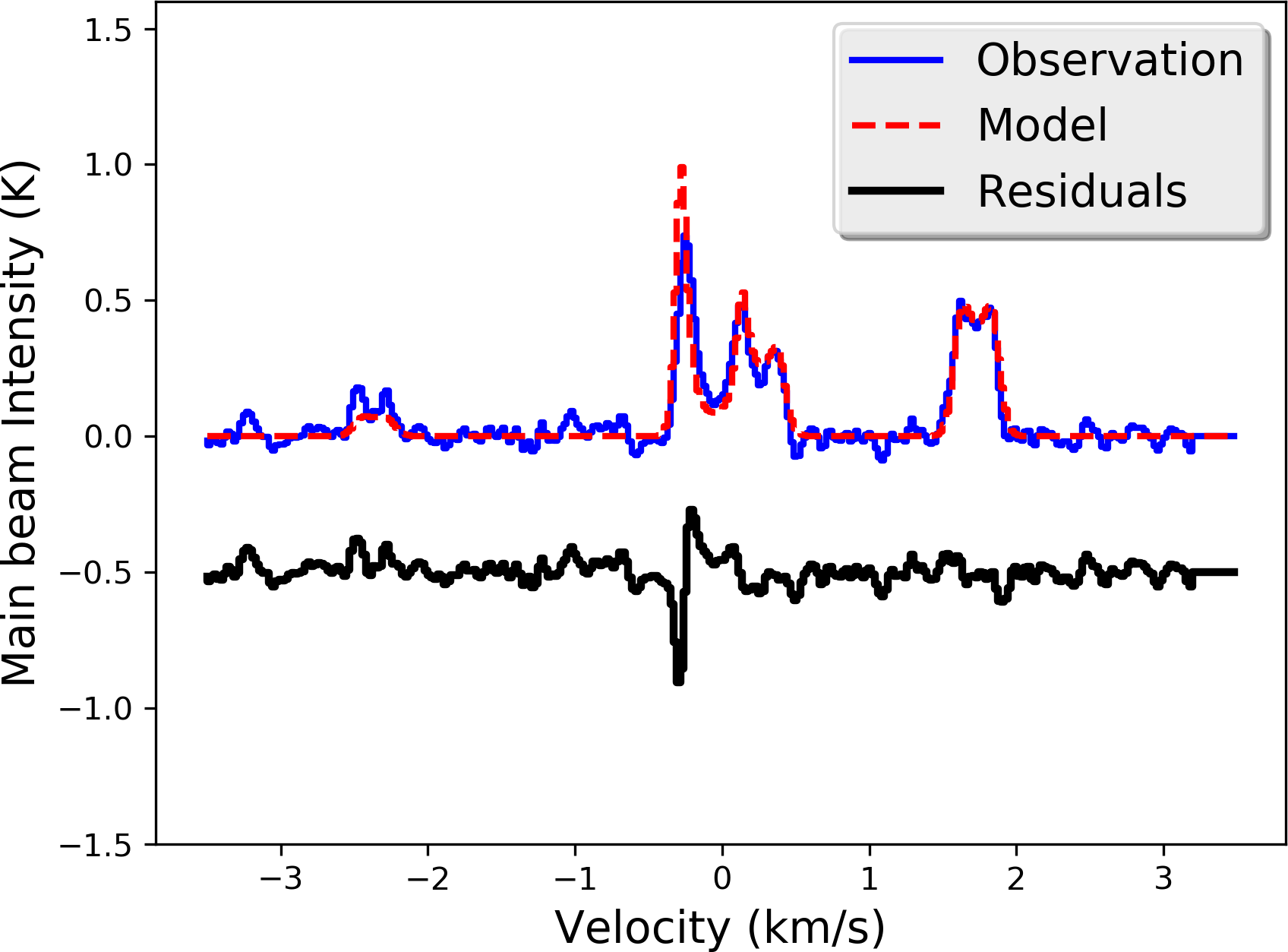}\\
  \caption{Line profiles of the \jtr{1}{0} (top row) and \jtr{3}{2}
    (bottom row) transitions of HCN computed with \texttt{ALICO} for
    different velocity profiles: static (left panels), uniform inward
    motion with $V(r)=-0.05\kms$(middle panels), and our best-fit
    profile from Eq.~\ref{eq:lesaffrevcollapse} (right panels).}
  \label{fig:hcnspecvelnovel}
\end{figure*}

In the static models, the HCN(3-2) line profiles are fairly well
reproduced, but the HCN(1-0) line profiles are not, with each
hyperfine component displaying double-peaked profiles. On the other
hand, models with a constant inward motion fail to reproduce the
features of HCN(3-2) line profiles.  The inward motions introduce red-blue asymmetries in the strong satellite hyperfine component at
1.8~km/s. They also make the strongest peak as twice stronger than the
observed value. Despite this, the constant inward-motion models are
able to reproduce the characteristics of the HCN(1-0) line profiles,
even though not perfectly. Last, models following the profile
described in equation \ref{eq:lesaffrevcollapse} can reproduce the
features of both the HCN(3-2) and the HCN(1-0) line profiles, as long as $r_V$
and $\Delta r_V$ are adequately chosen.

\subsection{Nonthermal velocity dispersion profile}

 \begin{figure*}

\includegraphics[width=\ww]{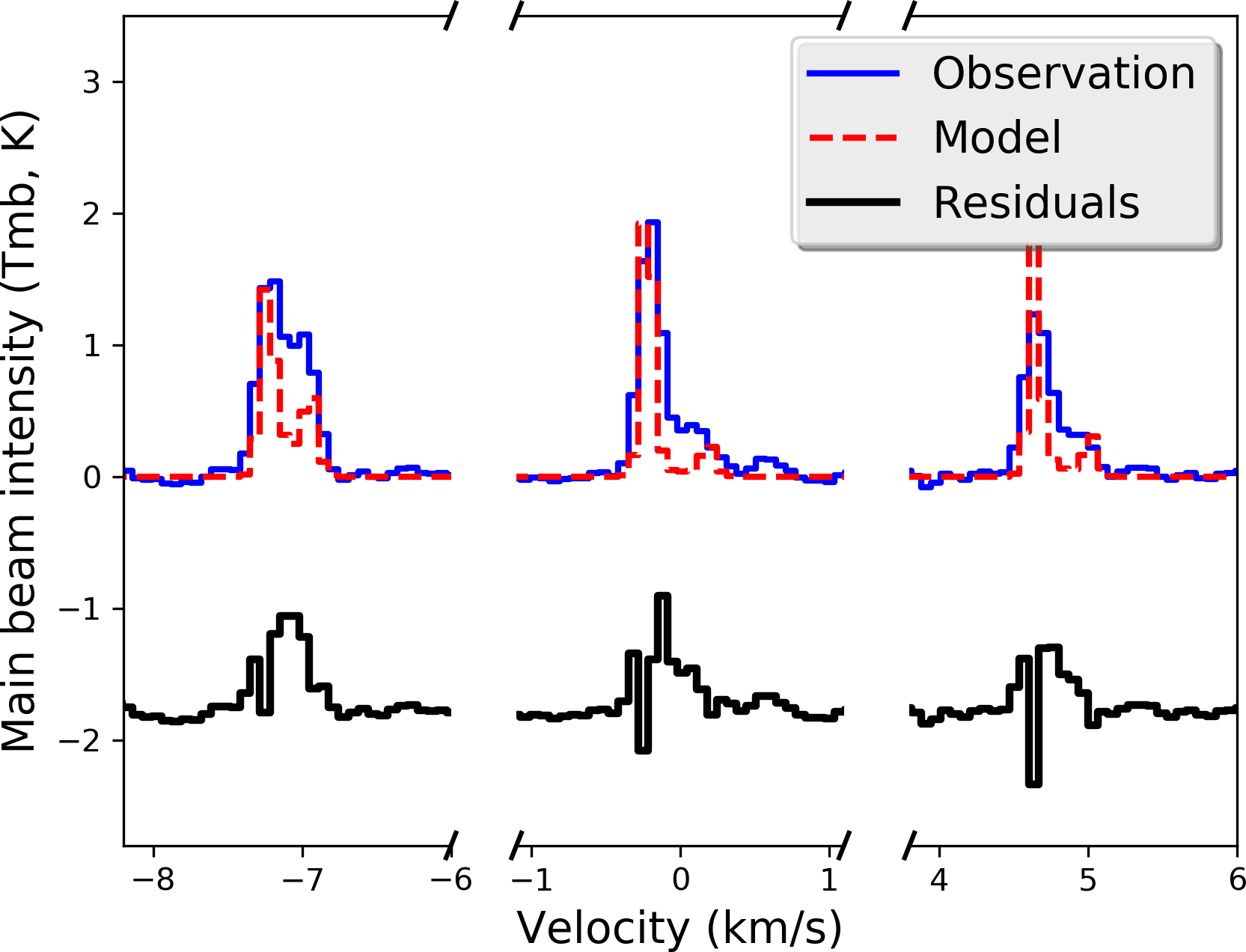}\hfill%
\includegraphics[width=\ww]{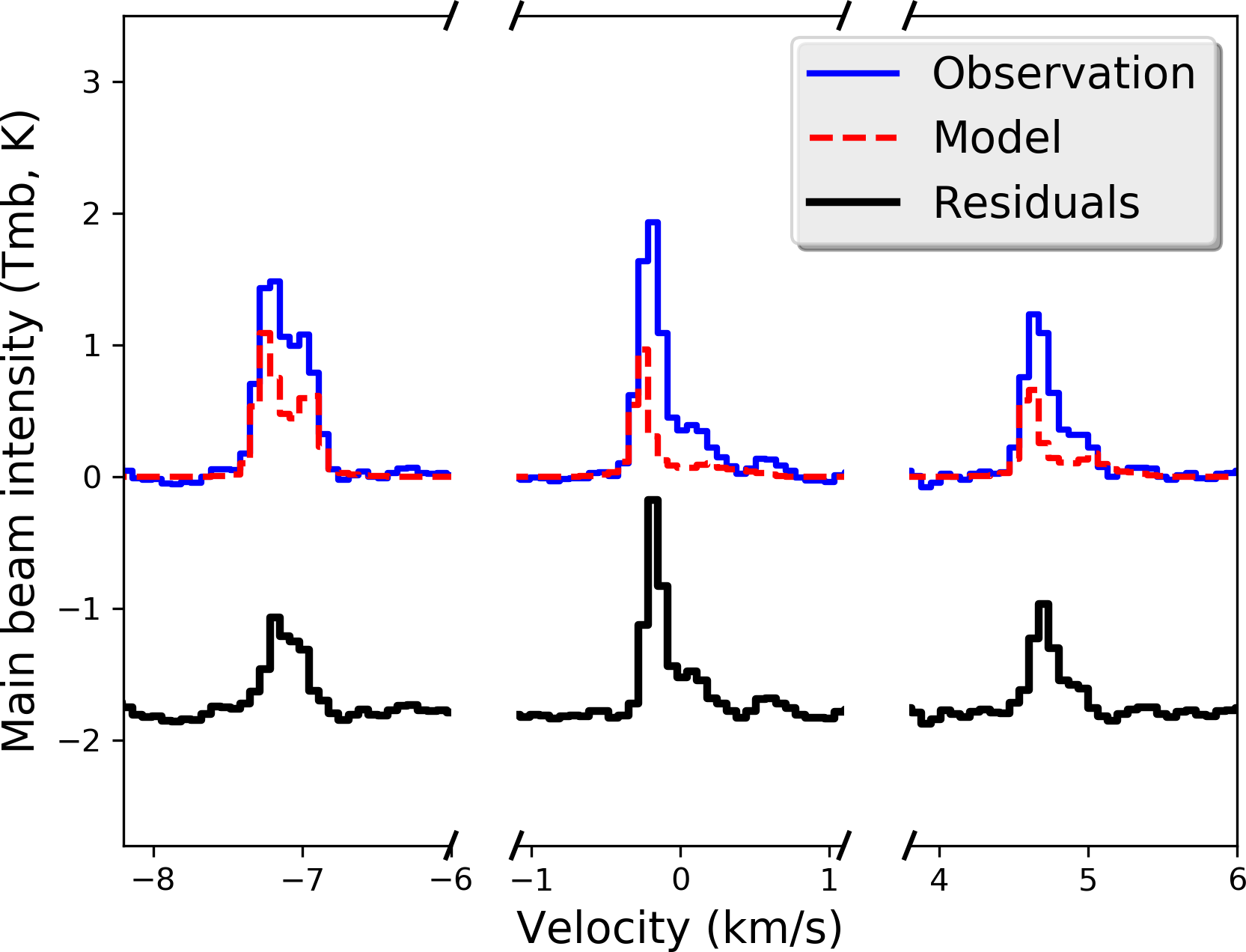}\hfill%
\includegraphics[width=\ww]{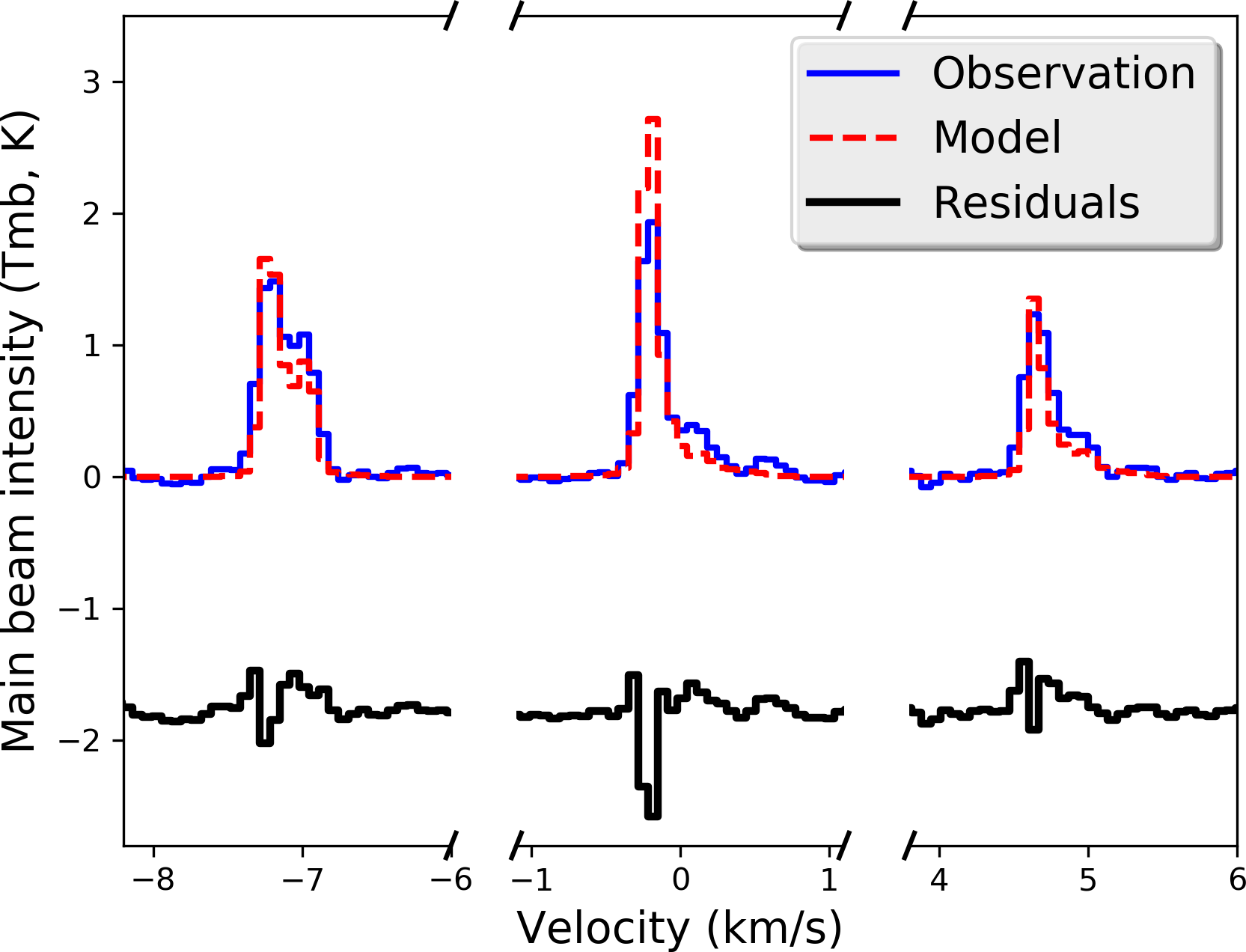}\\
\includegraphics[width=\ww]{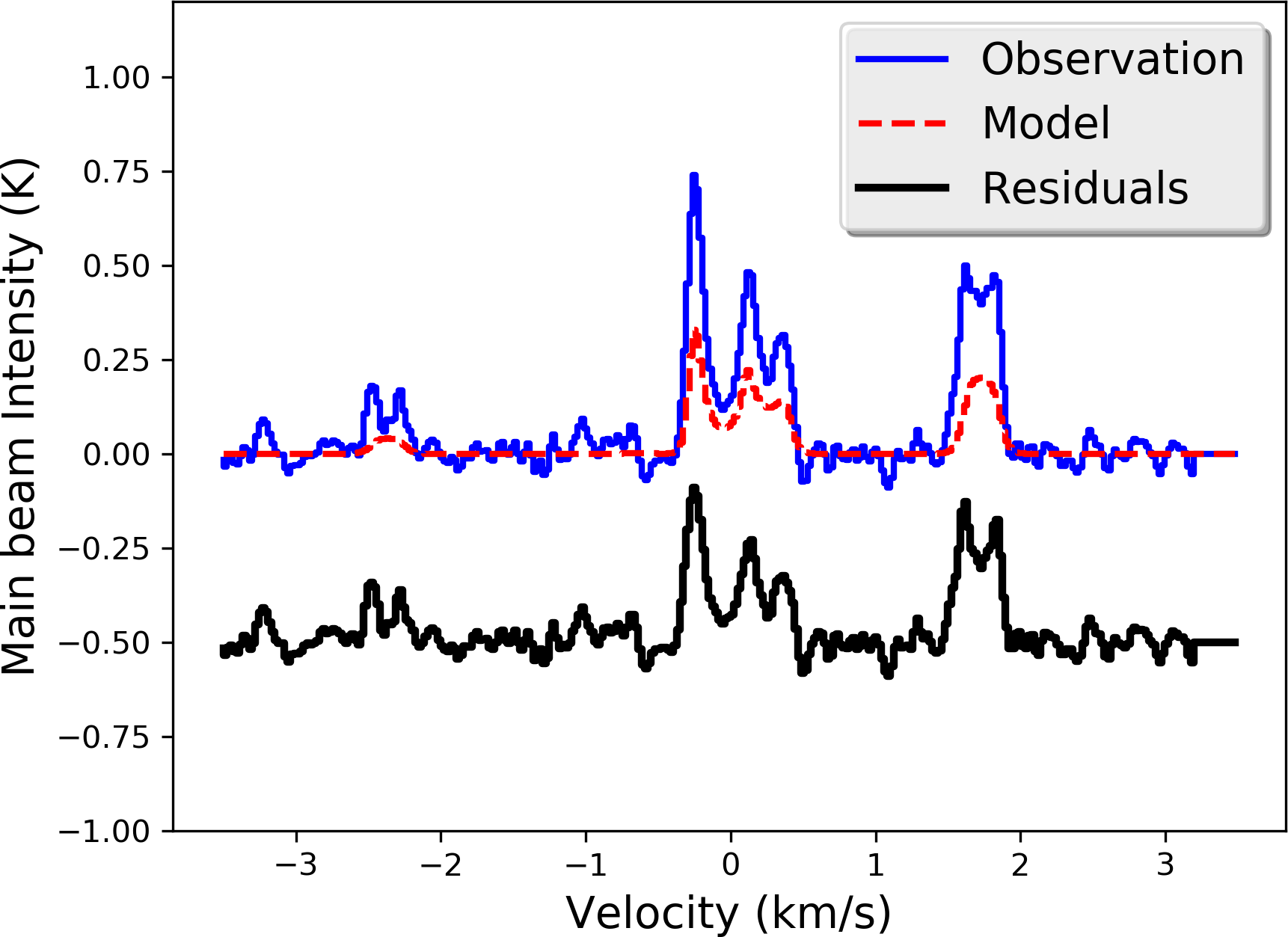}\hfill%
\includegraphics[width=\ww]{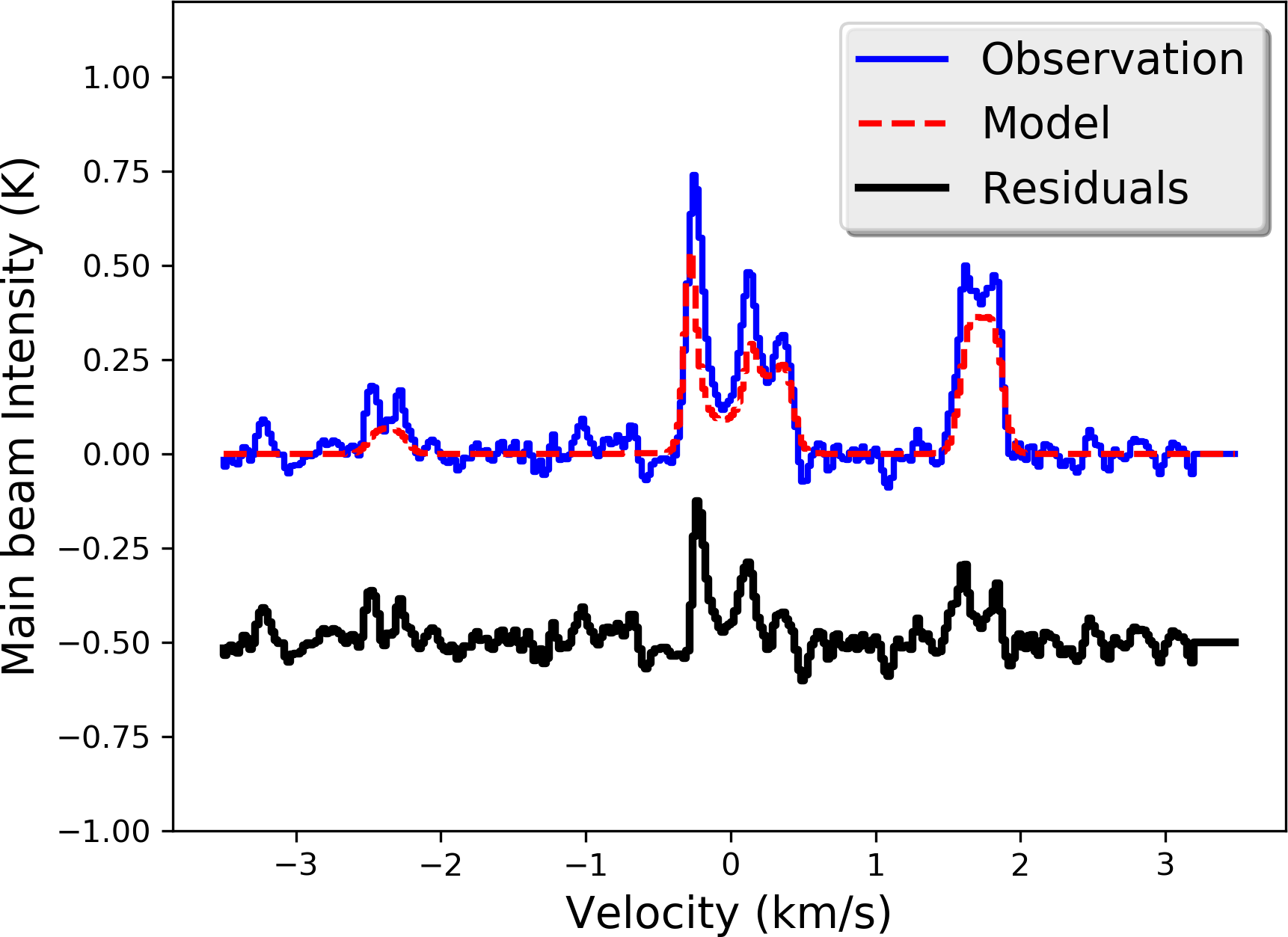}\hfill%
\includegraphics[width=\ww]{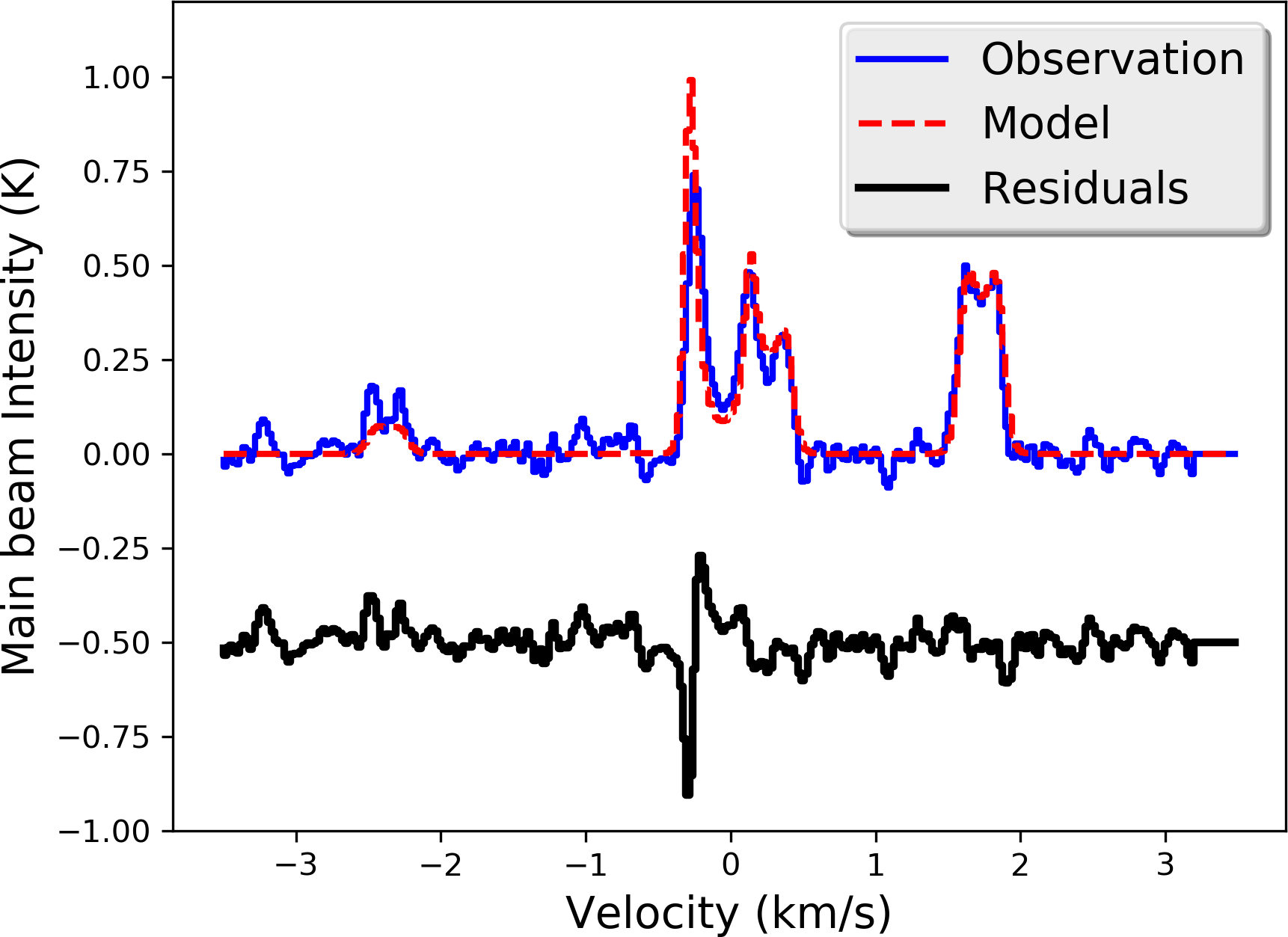}\\

\caption{Line profiles of the \jtr{1}{0} (top row) and \jtr{3}{2}
  (bottom row) transitions of HCN computed with \texttt{ALICO} for
  different nonthermal broadening profiles.  Constant
  $\sigma_{\mathrm{nth}}(r) = 0.0488$~km/s (left panels), linear
  increase with radius
  $\sigma_{\mathrm{nth}}(r) = 0.0488+(0.2805-0.0488)r$~km/s (middle
  panels), and our best-fit profile from Eq.~\ref{eq:dispersion} (right
  panels).}

\label{fig:nonthermalprofeffects}
\end{figure*}

 \begin{figure*}

   \includegraphics[width=\ww]{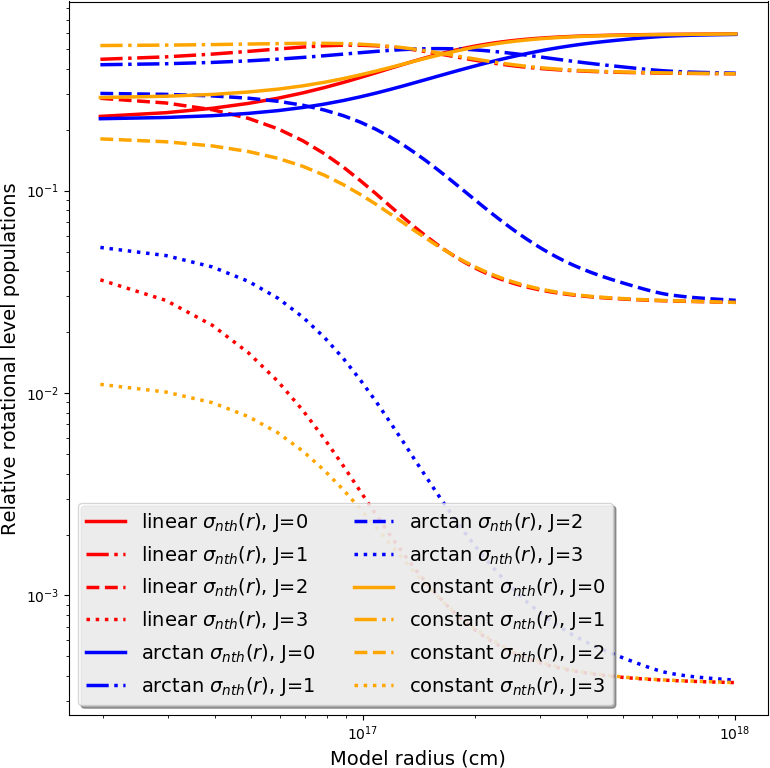}\hfill%
   \includegraphics[width=\ww]{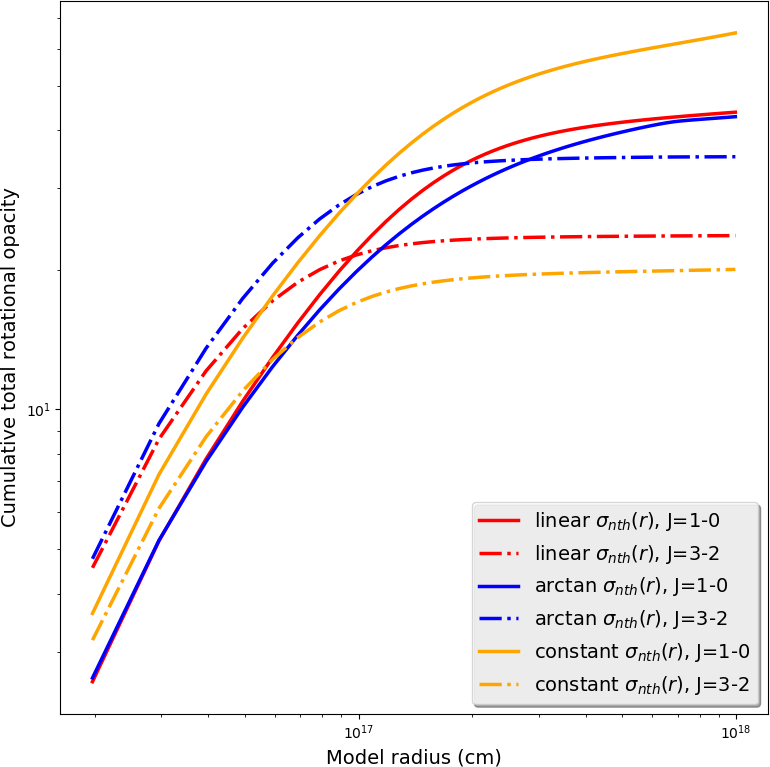}\hfill%
   \includegraphics[width=\ww]{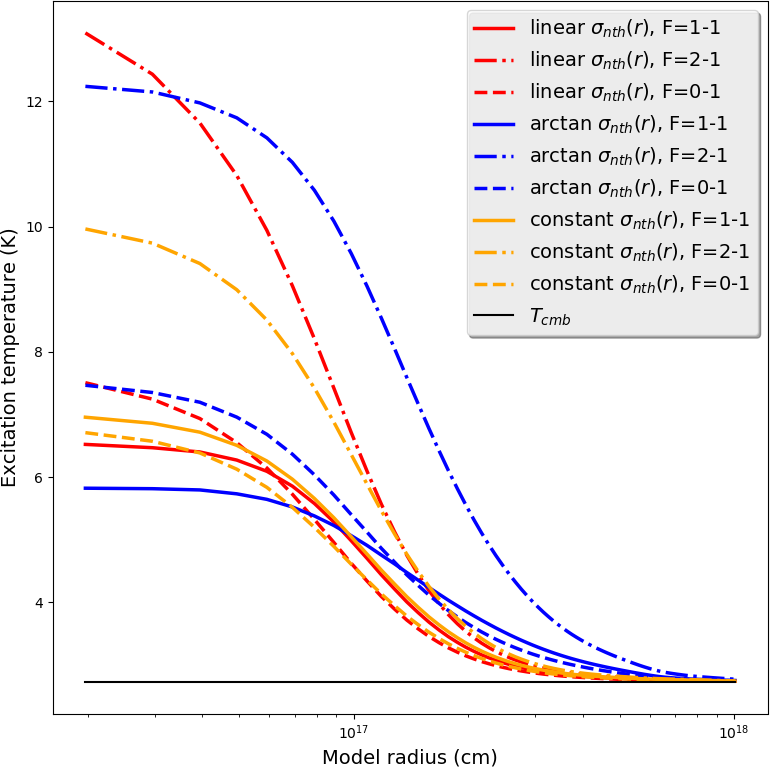}\\
   
   \caption{Relative rotational level populations (left),
     cumulative total rotational opacities (middle), and excitation
     temperatures (right) for the hyperfine components of the \jtr{1}{0}
     transition for the different $\sigma_{\mathrm{nth}}$
     profiles described in Figure
     \ref{fig:nonthermalprofeffects}. Blue lines correspond to the
     profile in eq. \ref{eq:dispersion}, red lines to a linear
     function for $\sigma_{\mathrm{nth}}$ profile, and the orange lines
     correspond to constant $\sigma_{\mathrm{nth}}$.}
   \label{fig:turbcomparisons}
   \end{figure*}

In panels a) and d) of figure \ref{fig:nonthermalprofeffects}, we show that the a constant $\sigma_{\mathrm{nth}}$ cannot reproduce the observed \jtr{1}{0} and \jtr{3}{2} line profiles well.
The \jtr{3}{2} hyperfine components are underproduced, and the \jtr{1}{0} hyperfine components have incorrect line profiles: they show a large self-absorption dip at V = 0 km/s, and the line ratios between the hyperfine components are not anomalous, that is, the line ratios between the hyperfine components are within the expected ratios (1/5 $\geqslant$ \ftr{0}{1}/\ftr{2}{1} $\leqslant$ 1 and 3/5 $\geqslant$ \ftr{1}{1}/\ftr{2}{1} $\leqslant$ 1).
On the other hand, a linear expression for $\sigma_{\mathrm{nth}}$ is not capable of reproducing the line profiles either.
The \jtr{3}{2} hyperfine components are slightly underproduced, while the \jtr{1}{0} hyperfine components  are severely underproduced and the line ratios between the hyperfine components are even more anomalous than the observed ratios (panels b) and e) ).
Finally, in panels c) and f), we show a good reproduction of the profiles of both transitions using the profile described in eq. \ref{eq:dispersion}.

The great differences seen between the different types of  $\sigma_{\mathrm{nth}}$ profiles can be understood when we take into into account the overlap of hyperfine components that occurs in \hcn\ rotational transitions with $J_{\mathrm{upper}}>1$ \citep{gottlieb1975,guilloteau1981,gonzalez-alfonso1993}.
Owing to the small line widths seen in L1498 (FWHM $\sim$ 0.2 km/s), under a constant and small $\sigma_{\mathrm{nth}}$ , there is very little overlap between the hyperfine components.
In this case, the excitation temperatures of the hyperfine components of the \jtr{1}{0} transition follow the expected trend, the highly optically thick \ftr{2}{1} component ($\tau \sim72$) is thermalized at the center of the model, with the excitation temperature decreasing in the outer layers, the other hyperfine components are less thick, ($\tau \sim14$ and $\tau\sim43$ for \ftr{0}{1} and \ftr{1}{1} respectively, and therefore farther from thermalization (Figure \ref{fig:turbcomparisons} panels b) and c) ). 
This scenario has no extra pumping of the population to $J=2$ and $J=3$ levels because of the radiative trapping caused by the hyperfine overlap, which decreases the \jtr{3}{2} excitation temperature (average \jtr{3}{2} \tex $\sim${3.6}~K) and in turn underproduces the line emission.

In the case of a linear increase of $\sigma_{\mathrm{nth}}$ with radius, we see that some radiative pumping of $J=2$ and $J=3$ is caused by the hyperfine component overlap, which causes  $J=2$ and $J=3$ to be more populated here than in the case of constant $\sigma_{\mathrm{nth}}$ (Figure \ref{fig:turbcomparisons} panel a) ).
This creates an excitation temperature imbalance in the \jtr{1}{0} transition at small radii, the \ftr{2}{1} component becomes supra-thermal,
and the excitation temperature of \ftr{1}{1} is lower than \ftr{0}{1}.
Because $\sigma_{\mathrm{nth}}$ continues to increase at a certain point, however, photons can easily escape in transitions \jtr{2}{1} and \jtr{3}{2,} leading to a fast depopulation of these levels and to a fast drop in excitation temperature of the \ftr{2}{1} component, which due to its high opacity ($\tau\sim$48, Figure \ref{fig:turbcomparisons} panel b) ) becomes heavily self-absorbed.
This explains the extreme hyperfine anomalies in panel b) of Figure \ref{fig:nonthermalprofeffects}.

The $\sigma_{\mathrm{nth}}$ profile described in eq. \ref{eq:dispersion} solves the problems presented by the other two profiles in a simple way.
The small increase in the nonthermal broadening with radii at small radii allows for the pumping of the $J=2$ and $J=3$ levels to proceed to larger radii, creating the observed \jtr{1}{0} hyperfine anomalies and keeping the \jtr{3}{2} excitation high enough to account for the observed intensities (average \jtr{3}{2} \tex$\sim$4.7~K).

\section{Spectral line fitting with \texttt{ALICO}}
\label{sec:alicomcmc}
\newcommand{\sumn}[1]{\ensuremath{\sum_{n_{#1}}}}

The models were evaluated by comparing the integrated intensity ($W$), peak intensity ($I$), and the velocity of the peak intensity ($V$) for multiple windows in each spectrum.
The windows were chosen in a way so as to contain only one peak.
Double-peaked lines were divided into two windows, whereas single-peaked lines were treated in a single window.
This approach was chosen to avoid the uncertainties of the channel-by-channel calculation of  $\chi^2$.
Since each channel is dependent on the surrounding channels because
of the line widths, channel-by-channel comparisons do not use independent samples.
The non-normalized $\chi^2$, $\chi^2_ {\mathrm{abs}}$ is then computed as\begin{equation}
\begin{split}
\chi^2_ {\mathrm{abs}} = \sumn{mol}\sumn{off}\sumn{win}\left[
\frac{(W_m-W_o)^2}{\sigma_{W}^2} + \frac{(I_m-I_o)^2}{\sigma_{I}^2} +\right.\\ 
&\left.+ \frac{(V_m-V_o)^2}{\sigma_{V}^2}\vphantom{} \right]
\end{split}
.\end{equation}
The $_m$ and $_o$ subscripts refer to the model and observations, respectively.
\(\sigma_{\mathrm{W}}\) is the uncertainty of the observed integrated intensity $W_o$,
\(\sigma_{\mathrm{I}}\) is the uncertainty of the observed peak intensity $I_o$ and,
\(\sigma_{\mathrm{V}}\) is the uncertainty of the observed velocity of the peak intensity $V_o$, the spectral resolution.
\sumn{win} is the sum over all
windows in a given spectrum, \sumn{off} is the sum over all offsets
that a molecule has been considered for the fitting procedure, and
finally, \sumn{mol} is the sum over all the molecules for which we executed the fit simultaneously.
$\chi^2_{\mathrm{abs}}$ is fed into the MCMC code so that it can
choose the parameters for the next set of walkers.

\end{document}